\documentclass[12pt]{article}

\usepackage[body={18cm, 21cm},right=1.8cm]{geometry}
\usepackage[english]{babel}
\usepackage{graphicx}
\usepackage{epstopdf}
\usepackage{amsmath}
\usepackage{amsfonts}
\usepackage{amssymb}
\usepackage{color}
\usepackage{mathrsfs}

\usepackage{array}
\usepackage{floatrow}
\newfloatcommand{capbtabbox}{table}[][\FBwidth]
\usepackage{caption}

\pdfoutput=1

\usepackage[colorlinks,bookmarks]{hyperref}
\definecolor{linkblue}{rgb}{0,0,0.8}
\definecolor{linkgreen}{rgb}{0,0.5,0}

\hypersetup{pdfpagemode=UseNone, pdfstartview=FitH, linkcolor=linkblue, %
            citecolor=linkgreen, urlcolor=linkblue}

\usepackage{graphicx}
\usepackage{epstopdf}
\usepackage{amsmath}
\usepackage{amsfonts}
\usepackage{amssymb}

\newcommand{\website}{\url{http://web.stanford.edu/~senatore/}}

\newcommand\nn{\nonumber}
\newcommand\eea{\end{eqnarray}}
\newcommand\bea{\begin{eqnarray}}
\newcommand{\sfrac}[2]{{\textstyle\frac{#1}{#2}}}

\def\beq{\begin{equation}}
\def\eeq{\end{equation}}
\def\d{\partial}

\newcommand{\be}{\begin{equation}}
\newcommand{\ee}{\end{equation}}
\newcommand{\ba}{\begin{align}}
\newcommand{\ea}{\end{align}}
\newcommand{\bg}{\begin{gather}}
\newcommand{\eg}{\end{gather}}
\newcommand{\bseq}{\begin{subequations}}
\newcommand{\eseq}{\end{subequations}}

\newcommand{\knl}{k_{\rm NL}}

\newcommand{\hinvMpc}{h\,$Mpc$^{-1}}

\newcommand{\q}{\vec{q}}

\definecolor{purple}{rgb}{0.78,0.18,0.77}

\definecolor{verde}{rgb}{0,0.5,0}

\def\epsdl{\epsilon_{\delta<}}

\def\xfl{{\vec x_{\rm fl}}}

\def\km{{k_{\rm M}}}

\makeatletter
\newlength{\apb@width}
\newcommand{\autoparbox}[2][c]{\settowidth{\apb@width}{#2}\parbox[#1]{\apb@width}{#2}}

\makeatother

\allowdisplaybreaks

\usepackage{ifthen}
\usepackage{amsmath}
\usepackage{amssymb}
\usepackage{color}
\usepackage{graphicx}

\renewcommand{\vec} {\textbf}
\newcommand{\V} {\textbf}
\newcommand*{\df}  {\delta}

\newcommand*{\non}  {\nonumber}

\def\xfl{{\textbf{x}_{\rm fl}}}

\def\km{{k_{\rm M}}}

\begin{document}

\vspace{5mm}
\vspace{0.5cm}
\begin{center}

\def\thefootnote{\fnsymbol{footnote}}
{\Large \bf Very Massive Tracers and Higher Derivative Biases}
\\[0.8cm]


{\large  Tomohiro Fujita${}^{1}$, Valentin Mauerhofer${}^{1,2}$, Leonardo Senatore${}^{1,3}$, \\[0.3cm]  Zvonimir Vlah${}^{1,3}$ and Raul Angulo${}^{4}$}
\\[0.5cm]

{\normalsize {${}^1$ \sl Stanford Institute for Theoretical Physics and Department of Physics, \\Stanford University, Stanford, CA 94306} }\\
\vspace{.3cm}

{\normalsize {${}^2$ \sl Institut de Th\'{e}orie des Ph\'{e}nom\`{e}nes Physique, \\ \'{E}cole Polytechnique F\'{e}d\'{e}rale de Lausanne, CH-1015 Lausanne, Switzerland}}\\
\vspace{.3cm}

{\normalsize {${}^3$ \sl Kavli Institute for Particle Astrophysics and Cosmology, \\ Stanford University and SLAC, Menlo Park, CA 94025}}\\
\vspace{.3cm}

{\normalsize {${}^4$ \sl Centro de Estudios de Fisica del Cosmos de Aragon,\\ Plaza San Juan 1, Planta-2, 44001, Teruel, Spain
}}

\vspace{.3cm}

\end{center}

\vspace{.8cm}

\hrule \vspace{0.3cm}
{\small  \noindent \textbf{Abstract} \\[0.3cm]
\noindent Most of the upcoming cosmological information will come from analyzing the clustering of the Large Scale Structures (LSS) of the universe through LSS or CMB observations. It is therefore essential to be able to understand their behavior with exquisite precision.  The Effective Field Theory of Large Scale Structures (EFTofLSS) provides a consistent framework to make predictions for LSS observables in the mildly non-linear regime. In this paper we focus on biased tracers. We argue that in calculations at a given order in the dark matter perturbations, highly biased tracers will underperform because of their larger higher derivative biases. A natural prediction of the EFTofLSS is therefore that by simply adding higher derivative biases, all tracers should perform comparably well. We implement this prediction for the halo-halo and the halo-matter power spectra at one loop, and the halo-halo-halo, halo-halo-matter, and halo-matter-matter bispectra at tree-level, and compare with simulations. We find good agreement with the prediction: for all tracers, we are able to match the bispectra up to $k\simeq0.17\hinvMpc$ at $z=0$ and the power spectra to a higher wavenumber.
}

 \vspace{0.3cm}
\hrule
\def\thefootnote{\arabic{footnote}}
\setcounter{footnote}{0}

\vspace{.8cm}

\newpage
\tableofcontents




\section{Introduction and main ideas}
\label{sec:intro}

Current and next generation CMB and galaxy surveys, such as eBOSS~\cite{Dawson:2015wdb}, LSST~\cite{Abell:2009aa}, DESI~\cite{Levi:2013gra} and Euclid~\cite{Refregier:2010ss}, SPT~\cite{Ruhl:2004kv} and ACT~\cite{Thornton:2016wjq}, will measure the statistical distribution of  cosmological large-scale structures with percent/sub-percent precision~\cite{Percival:2013awa}.  In order to fully exploit these cosmological data, it is important to be able to have a way to make theoretical predictions with comparable or better accuracy.

While in the last couple of decades numerical simulations have been the main tool to predict the clustering of large scale structures, in the last few years the advent of the so-called Effective Field Theory of Large Scale Structures (EFTofLSS)~\cite{Baumann:2010tm,Carrasco:2012cv,Porto:2013qua,Senatore:2014via} has allowed the development of an analytic approach that is able to predict the Large Scale Structure (LSS) correlation functions with exquisite precision in the so-called mildly non-linear regime, where density fluctuations are still safely smaller than one~\cite{Baumann:2010tm,Carrasco:2012cv,Porto:2013qua,Senatore:2014via,Carrasco:2013sva,Carrasco:2013mua,Pajer:2013jj,Carroll:2013oxa,Mercolli:2013bsa,Angulo:2014tfa,Baldauf:2014qfa,Senatore:2014eva,Senatore:2014vja,Lewandowski:2014rca,Mirbabayi:2014zca,Foreman:2015uva,Angulo:2015eqa,McQuinn:2015tva,Assassi:2015jqa,Baldauf:2015tla,Baldauf:2015xfa,Foreman:2015lca,Baldauf:2015aha,Baldauf:2015zga,Bertolini:2015fya,Bertolini:2016bmt,Assassi:2015fma,Lewandowski:2015ziq,Cataneo:2016suz,Bertolini:2016hxg}. In particular, the EFTofLSS has been applied to the description of the dark matter two-point function~\cite{Carrasco:2012cv,Senatore:2014via,Carrasco:2013mua,Foreman:2015lca,Baldauf:2015aha}, three-point function~\cite{Angulo:2014tfa,Baldauf:2014qfa}, four-point function (which includes the covariance of the power spectrum)~\cite{Bertolini:2015fya,Bertolini:2016bmt}; to the dark matter momentum power spectrum~\cite{Senatore:2014via,Baldauf:2015aha}, to the displacement field~\cite{Baldauf:2014qfa}; and to the vorticity slope~\cite{Carrasco:2013mua,Hahn:2014lca}. The effects of baryons on the power spectrum have been incorporated in~\cite{Lewandowski:2014rca}.  The extension of the EFTofLSS to biased tracers has been carried out in~\cite{Senatore:2014vja}, and the predictions compared to data for the power spectrum and bispectrum in~\cite{Angulo:2015eqa}. Redshift space distortions~\cite{Senatore:2014vja,Lewandowski:2015ziq}, and the impact of primordial non-Gaussianity on large scale structure observables~\cite{Angulo:2015eqa,Assassi:2015jqa,Assassi:2015fma,Lewandowski:2015ziq} have also been recently included. Fast implementations of the predictions of the EFTofLSS to efficiently explore their dependence on various cosmological parameters have been recently developed in~\cite{Cataneo:2016suz}, with public codes available at the following website~\footnote{\website}.

This paper will focus on the EFTofLSS when applied to biased tracers, such as halos or galaxies. In particular, we will focus on tracers which are highly massive, and therefore highly biased. In the EFTofLSS biased tracers are represented as a functional of the second derivatives of the  gravitational fields, of matter fields, such as dark matter (denoted by the subscript ${}_c$) or baryons (denoted by the subscript ${}_b$), of stochastic terms $\epsilon$, and of spatial derivatives, as well as of the parameters of the background cosmology, such as $\Omega_{dm}$, as well as of all other parameters that determine the laws of nature, such as the electron mass $m_e$, and in general of all terms allowed by general covariance. All of these terms need to be evaluated on the past light cone of the spacetime point of interest. Schematically, we have the tremendous expression~\cite{Senatore:2014vja}
\be\label{eq:euler_bias_0}
\delta_M(\vec x,t)= f\left(\left.\{\d_i\d_j \phi(\vec x',t'), \delta_b, \d_j v_c^i(\vec x',t'), \frac{\d^i}{\km},\epsilon(\vec x',t'), \ldots,\Omega_c, \ldots, m_e,\ldots\}\right|_{\rm on \ past\ lightcone}\right)\ ,
\ee

If we are interested in spatial fluctuations of this quantity, we realize that only the fluctuating fields in (\ref{eq:euler_bias_0}) carry spatial dependence. If we are interested in long wavelength perturbations, the fluctuations are small, and we can Taylor expand (\ref{eq:euler_bias_0}) to drastically simplify it and obtain, schematically,
\bea\label{eq:euler_bias_2}
&&\delta_M(\vec x,t)\simeq \int^t dt'\; H(t')\; \left[  \sum_{j=c,b}  \bar c_{\d^2\phi,j}(t,t')\; \delta_j(\xfl,t') \right.\\  \nonumber
&&\quad+\sum_{j=c,b} \bar c_{\d_i v^i,j}(t,t') \;  \frac{\d_i v^i_j(\xfl,t')}{H(t')}+\bar c_{\d_i \d_j \phi \d^i \d^j \phi}(t,t') \;\frac{\d_i\d_j \phi(\xfl, t')}{H(t')^2}\frac{ \d^i \d^j \phi(\xfl,t')}{H(t')^2} + \ldots\\\nonumber
&&\quad+ \bar c_{\epsilon}(t,t')\;\epsilon(\xfl,t')+\bar c_{\epsilon\d^2\phi}(t,t') \;\epsilon(\xfl,t')\frac{\d^2\phi(\xfl,t')}{H(t')^2}+ \ldots \\ \nonumber
&&\left.\quad+  \bar c_{\d^4\phi}(t,t')   \;\frac{\d^2_{x_{\rm fl}}}{\km^2}\frac{\d^2\phi(\xfl,t')}{H(t')^2}+\dots\ \right] .
\eea
Here $\bar c_{\ldots}(t,t')$ are dimensionless kernels with support of order one Hubble time and with size of order one, $\xfl$ represents the location at time $t'$ of the fluid element that at time $t$ is at location ${\bf x}$, and the scale~$\km$ here is the comoving wavenumber enclosing the mass of an object~\cite{Senatore:2014eva} (we defer to later in the text for more explicit definitions).

How to include the effect of baryons for biased tracers was introduced in~\cite{Angulo:2015eqa,Schmidt:2016coo}, which of course required first to understand that baryons can be treated as an effective fluid-like system similar to dark matter, which was done in~\cite{Lewandowski:2014rca}. 
In the presence of primordial non-Gaussianities~\cite{Angulo:2015eqa,Assassi:2015jqa,Assassi:2015fma,Lewandowski:2015ziq} the tracer fields depend on additional fields, $\tilde\phi(\xfl(t,t_{\rm in}),t_{\rm in})^{i_1,\ldots ,i_n}$, that can be formed out of the gravitational field, multiplied by some power $\alpha$ of the long wavenumber of interest, $k_L^\alpha$, with $0\leq \alpha\leq 2$, and potentially by some additional factor associated to the angle of $\vec k_L$, and then divided by the transfer function $T(k)$ of the primordial fluctuations. Notice also the peculiar value of the coordinates $(\xfl(t,t_{\rm in}),t_{\rm in})$  at which this field needs to be evaluated. We will neglect the effect of baryons and primordial non-Gaussianities for the rest of this paper, though all what we describe can be trivially extended to include these cases.

The time integrals over unknown kernels that appear in~(\ref{eq:euler_bias_2}) might make it seem not a very useful expression. However, the structure of the perturbative solutions comes to our help. In perturbation theory, the solution at a given order is the sum of products of a function of time, approximately equal to the a power of the linear growth factor, times a function of wavenumber. Therefore, by plugging in (\ref{eq:euler_bias_2}) the perturbative solution, we can formally evaluate the time integrals to obtain an expression where each term in perturbation theory is multiplied by his own bias. Schematically, we have~\cite{Senatore:2014vja}
\bea\label{eq:euler_bias_4_intro}
&& \delta_h(\bold{k}, t) = \nonumber
\\
&& \quad = c_{\delta, 1}(t) \delta^{(1)}(\bold{k}, t) + c_{\delta, 2}(t) \delta^{(2)}(\bold{k}, t) + c_{\delta, 3}(t) \delta^{(3)}(\bold{k}, t) + c_{\delta, 3_{c_s}}(t) \delta^{(3)}_{c_s}(\bold{k}, t) +\ldots \nonumber
\\ \nn
&& \qquad + [c_{\delta, 1}(t) - c_{\delta, 2}(t)] \left[\partial_i \delta^{(1)} \frac{\partial^i}{\partial^2} \theta^{(1)} \right]_{\bold{k}}(t) + [c_{\delta, 2}(t) - c_{\delta, 3}(t)] \left[\partial_i \delta^{(2)} \frac{\partial^i}{\partial^2} \theta^{(1)} \right]_{\bold{k}}(t) +\ldots\\ 
&& \qquad+  \bar c_{\d^4\phi}(t,t')   \;\frac{\d^2_{x}}{\km^2}\delta ^{(1)}+\dots\  .
\eea

After the renormalization is performed, the loop expansion, completed by the insertion of the relevant higher order bias coefficients, amounts to an expansion in the parameters that control the dark matter expansion: $\epsilon_{\delta<}$ and $\epsilon_{s>}$~\cite{Porto:2013qua,Senatore:2014via}. These are defined as 
\bea
&&\epsilon_{s >} =k^2  \int_k^\infty \frac{d^3k' }{ (2 \pi)^3}  \frac{P_{11}(k') }{ k'^2}\ , \qquad \epsilon_{\delta <} = \int^k_0 \frac{d^3k' }{ (2 \pi)^3} P_{11}(k')\ ,
\eea
where $P_{11}(k)$ is the dark matter power spectrum. $\epsilon_{s >} $ represents the displacement due to short wavelength modes, while $\epsilon_{\delta <}$ represents the tidal force due to long wavelength modes. Both of these scale  proportionally to $k/\knl$. For simplicity, we will refer to $\epsilon_{\delta<}$ and $\epsilon_{s>}$ with the common symbol of $\epsilon_{\delta}$.  In the Eulerian treatment we expand also in displacement due to long wavelength modes $\epsilon_{s<}= (k\, \delta s_<)^2$, where~\footnote{For IR-safe quantities, the relevant parameters is~\cite{Senatore:2014via} 
\be\label{eq:epssafe}
\epsilon_{s_<}^{\rm safe} =k^2  \int_{k_{\rm bao}}^k  \frac{d^3k' }{ (2 \pi)^3}  \frac{P_{11}(k') }{ k'^2}\ ,
\ee
where $k_{\rm bao}$ is the wavenumber associated to the inverse of the bao peak length.
}
\bea
\epsilon_{s_<} &=&k^2  \int_0^k \frac{d^3k' }{ (2 \pi)^3}  \frac{P_{11}(k') }{ k'^2}\ .
\eea
As described in~\cite{Senatore:2014via}, $\epsilon_{s<}$ is of order one for the $k$'s of interest, and therefore one cannot Taylor expand in this parameter. However, as explained in~\cite{Senatore:2014via,Angulo:2014tfa,Lewandowski:2014rca} for dark matter and baryons, and in~\cite{Senatore:2014vja,Angulo:2015eqa} for redshift space distorsions, one can resum exactly in this parameter. Instead, the expansion in higher derivative bias coefficients corresponds to an expansion in 
\be
\epsilon_M\sim\left(\frac{k}{\km}\right)^2\ .
\ee
Finally, the expansion in stochastic bias terms offers yet another parametric dependence, since $\langle\epsilon^2\rangle\sim \frac{1}{\bar n_M}$, with $\bar n_M$ being the number density of the population of biased tracers under consideration.

Schematically, we therefore have the following perturbative expansion for biased tracers~\cite{Senatore:2014vja,Angulo:2015eqa}
\bea\label{eq:dark_galaxy_galaxy_expansion}
&&\langle\delta_M(k)\delta_M(k)\rangle'\sim c_{\delta} \left\{  \underbrace{\left[1+\left(\frac{k}{\km}\right)^2+\ldots\right]}_\text{Bias Derivative Expansion: $k/k_M$}\times \underbrace{\left[1+\epsilon_{\delta<}+\ldots\right]}_\text{Matter Loop Expansions: $\epsilon_{\delta}$}\times \ P_{11}(k) \right\}
\\ \nonumber
&&\left. +\underbrace{\left[1+\left(\frac{k}{\km}\right)^2+\ldots\right]}_\text{Stochastic Bias Derivative Expansion: $k/k_M$}\times  \underbrace{c_{\delta}\left[1+\epsilon_{\delta<}+\ldots\right]}_\text{Mixed Matter Stochastic Bias Expansion: $\epsilon_{\delta}$}\times \underbrace{\frac{1}{\bar n_M}}_\text{Stochastic Bias: $1/\bar n_M$}  \right. \ .
\eea
Of the above perturbative expression, we still need to explain how the bias coefficients appear, which was not specified in~\cite{Senatore:2014vja,Angulo:2015eqa}. We have put only one common factor in front of the expansion that is not proportional to the stochastic terms. Often in the community, it is assumed that the bias coefficients  $c_n$, such as $c_{\delta^n}$, multiplying terms of order $(\delta^{(1)})^n$ in~(\ref{eq:euler_bias_4_intro}) go as $c_{n}\sim (c_{1})^n$. Instead, as we describe in this paper, we find neither very strong justification nor evidence in the fitting to the data of such a behavior. We rather find that all bias coefficients are of comparable order, as expressed in~(\ref{eq:dark_galaxy_galaxy_expansion}): $c_{n}\sim c_{1}$.

The consistent perturbative expansion of (\ref{eq:dark_galaxy_galaxy_expansion}) was compared with simulation data on several statistic of halos in~\cite{Angulo:2015eqa}. In particular, in~\cite{Angulo:2015eqa} measurements of the halo-halo and halo-matter two point functions, and of the matter-matter-halo, the matter-halo-halo and halo-halo-halo three point functions for three different mass populations were matched to the predictions of the EFTofLSS. There, it was found that the EFTofLSS allows for a much improved match between theory and simulations. However, in the same paper~\cite{Angulo:2015eqa} it was found that the predictions for tracers characterized by a smaller mass performed better than the ones for more massive tracers. A look at expression~(\ref{eq:dark_galaxy_galaxy_expansion}) easily explains this fact. In fact, predictions in~\cite{Angulo:2015eqa} were made at one loop for the two-point functions and at tree level for the three-point functions. Performing the calculations at the same loop order for all the tracers means performing the calculation at the same order in $\epsdl$ for all the tracers. Since the more massive is the tracers, the smaller is $k_M$, if one does not include higher derivative terms for more massive tracers, then the predictions for these more massive tracers are doomed to fail at lower wavenumber. Viceversa, a prediction of the perturbative expansion in the EFTofLSS in eq.~(\ref{eq:dark_galaxy_galaxy_expansion}) is that, given a calculation at a given loop order, by solely adding higher derivative terms, the predictions for the tracers of all masses should fail approximately at the same wavenumber, when the common theoretical error from the next order in $\epsdl$ is the dominant error for all the tracers. The only reason for failing at different wavenumbers is the fact that there is an additional source of theoretical error that comes from the fact that the size of the bias coefficient might depend on the tracer. Once the error due to $k/k_M$ is made negligible, the theoretical error scales as $c (\epsilon_{\delta})^m \sim c (k/\knl)^{m(3+n)}$, where $m$ is the perturbative order of the calculation, and we took $\epsilon_\delta\sim (k/\knl)^{3+n}$, with $n\sim -1.7$ being that approximate slope of the power spectrum at the $k$ of interest. Therefore, we have that, if we threshold on a given error $\epsilon_{\rm error}$, this occurs at wavenumber $k \propto c^{-\frac{1}{m(3+n)}} $. In this expression, we see that the dependence on the tracer is simply relegated to the factor $c^{-\frac{1}{m(3+n)}}$, which, as the perturbative order $m$ in made higher, becomes smaller and smaller.

The purpose of this paper is to check the correctness of this prediction. First, we will add higher derivative operators for the bins associated to more massive halos, and we will indeed find that the predictions now fail approximately at the same wavenumber for all tracers. We have that the power spectra in \cite{Angulo:2015eqa} were computed to one-loop order, which correspond to order $\epsilon_{\delta} + \epsilon_M \epsilon_{\delta} + \epsilon_{\delta}^2$. The $\epsilon_M \epsilon_{\delta}$ part comes from the leading linear higher derivative term $\frac{\d^2}{k_M^2} \delta$, which was already included in \cite{Angulo:2015eqa}. The next leading term in $\epsilon_M$  scale as $\epsilon_M^2 \epsilon_{\delta}$, which is potentially larger than $\epsilon_M \epsilon_{\delta}^2$. The order of the terms we will add to the power spectra will therefore be $\epsilon_M^2 \epsilon_{\delta}$. Instead,  the bispectra were computed at tree level with no higher derivative terms. This corresponds to order $\epsilon_{\delta}^2$. We will therefore limit ourselves to add the higher derivative terms that contribute to order $\epsilon_M \epsilon_{\delta}^2$.

We can estimate the importance of the higher derivative terms by comparing the two expansion parameters $\left( k/k_M \right)^2$ and $\left( k/k_{\rm NL} \right)^{3+n}$ at different values of $k$. To do so, we use the approximate values $k_{\rm NL} \approx 4.6 \, h \, {\rm Mpc}^{-1}$ and $n \approx -1.7$, which are valid for a regime where $k \lesssim 0.25 \, h \, {\rm Mpc}^{-1}$, \cite{Carrasco:2013mua}. For~$k_M$, we use the rough estimate $k_M \sim 2 \pi \left( \frac{4 \pi}{3} \frac{\rho_{b,0}}{M_{\rm halo}} \right)^{1/3}$, where $\rho_{b,0} \simeq 2.6 \cdot 10^{-24} {\rm g}/{\rm m^3}$ is the background density of the universe. Notice that $k_M$ depends on the mass of the halo, that is why the higher derivative terms are important to predict the clustering of very high mass halos. In this work, the halos are separated into four bins, (Bin0, Bin1, Bin2, Bin3), according to their mass, from the lightest to the heaviest. The mass of the halos for each bin is given in \cite{Okumura:2012xh}. Table \ref{tb:kmvalue} presents the estimates for $k_M$ and the ratios $\left( k/k_M \right)^2$ for two different values of $k$ for each bin. These estimates are very rough, and should be taken at the order of magnitude level, but they already highlight the fact that the higher derivative terms are far more important for Bin2 and Bin3 than for Bin0 and Bin1, which explained why Bin2 was not matching the data up to the same maximum wavenumber $k_{\rm max}$ as Bin0 and Bin1 in \cite{Angulo:2015eqa}. Also we see that for Bin2 and Bin3 the order of magnitude of the higher derivative terms is comparable with the one of the perturbative expansion. Though very rough, these estimates encourage us to add the higher derivative terms.

\begin{table*}[t!]
\caption{The first line presents approximate values of $k_M$ for each mass bin. To compute those values, we use $k_M \sim 2 \pi \left( \frac{4 \pi}{3} \frac{\rho_{b,0}}{M} \right)^{1/3}$, with $M$ being the mass of the bin. The numerical values of the mass of the bins are given in \cite{Okumura:2012xh}. $\rho_{b,0}$ is the mean matter density in the universe, whose numerical value is around $\rho_{b,0} \simeq 2.6 \cdot 10^{-24} {\rm g}/{\rm m^3}$. The second and third lines are estimation of the ratio $\left(k/k_M\right)^2$ for two different values of $k$, namely $k_1 = 0.1 \, h \, {\rm Mpc}^{-1}$ and $k_2 = 0.15 \, h \, {\rm Mpc}^{-1}$.  We compare these values with the expansion parameter of the loops, $\epsdl\sim \left(k/k_{\rm NL} \right)^{3+n}$, where $k_{\rm NL} \approx 4.6 \, h \, {\rm Mpc}^{-1}$ and $n \approx -1.7$. We have $\left( k_1/k_{\rm NL} \right)^{3+n} \approx 7 \cdot 10^{-3}$ and $\left( k_2/k_{\rm NL} \right)^{3+n} \approx 1 \cdot 10^{-2}$.}
\centering 
\setlength{\tabcolsep}{8pt}
\renewcommand{\arraystretch}{1.0}
\begin{tabular}{c|cccccc}
\hline\hline
 & Bin0 & Bin1 & Bin2 & Bin3 \\ [0.5ex] 
\hline
$k_M [h \, \rm{Mpc}^{-1}]$  & $3.3-4.8$ & $2.3-3.3$ & $1.6-2.3$ & $1.1-1.6$ \\
$(k_1/k_M)^2$ & $(4.3-9.2) \cdot 10^{-4}$ & $(0.9-1.9) \cdot 10^{-3}$ & $(1.9-3.9) \cdot 10^{-3}$ & $(3.9-8.3) \cdot 10^{-3}$\\
$(k_2/k_M)^2$ & $(1.0-2.1) \cdot 10^{-3}$ & $(2.1-4.3) \cdot 10^{-3}$ & $(4.3-8.8) \cdot 10^{-3}$ & $(0.9-1.9) \cdot 10^{-2}$\\
\hline
\end{tabular}
\label{tb:kmvalue}
\end{table*}

In performing our study, we will find two additional ways to improve the findings of~\cite{Angulo:2015eqa}. 
 First, we will find a factor of two of mistake in a contribution to the prediction for the halo-halo-halo bispectra in~\cite{Angulo:2015eqa}~\footnote{The mistake was in the Mathematica notebook, not in the text. We apologize.}. After correcting for this factor of two, we find that the predictions of the EFTofLSS, even before adding the additional higher derivative biases, match much better the measurements in simulations. This is interesting because it offers yet another verification of the correctness of the EFTofLSS and of its predicting power. In the EFTofLSS there are free parameters, but there are also contributions that do not depend on these parameters, which are called calculable terms. The fact that if we make a mistake in the calculable terms we cannot match the data as well as when we compute these terms correctly, is proof that the functional freedom induced by the free parameters of the EFTofLSS is not strong enough to erase the contribution from the calculable terms. This result is therefore a statement of the correctness and of the predicting power of the EFTofLSS, notwithstanding the presence of free parameters.     
 
A last improvement with respect to~\cite{Angulo:2015eqa} concerns the way the predictions of the EFTofLSS are compared to simulation data. This procedure is delicate for two different reasons. First, the predictions of the EFTofLSS depend on parameters that need to be measured from the same set of data that are used to asses the accuracy of the EFT predictions. Since simulation data have smaller sampling variance at higher wavenumbers, we would like to measure them at high wavenumber. But, as it is evident from  (\ref{eq:dark_galaxy_galaxy_expansion}), the  inaccuracy of the EFT predictions grows as we move to higher wavenumber, which pushes us to measure these parameters at low wavenumber. We address this counteractive trends by implementing a fitting procedure very similar to the one developed in~\cite{Foreman:2015lca} that ensures that, as we move our fitting to higher and higher wavenumbers, we do not degrade the fit at lower wavenumbers (where our prediction is more accurate).

In Sec.~\ref{sec:equations} we construct the predictions for the two-point and three-point functions of highly massive tracers, and in Sec.~\ref{sec:results} we perform the comparison with simulations to determine the bias parameters and the accuracy of the predictions. We find that indeed higher mass bins match the data to a comparable level as the low mass bins, after the addition of the higher derivative terms.

\section{Biased objects in the EFTofLSS}
\label{sec:equations}

\subsection{Overdensity of biased objects}
\label{subsec:field}

The logic and structure of the computation of correlation functions of biased tracers in the EFTofLSS was explained in detailed in \cite{Senatore:2014eva}, and explicitly computed and compared to simulations for the first time in~\cite{Angulo:2015eqa}. The formulas that we present in this section are very similar to the ones of~\cite{Angulo:2015eqa}, apart for the higher derivative bias terms that we now include in the computation. We will therefore be quite schematic and refer to~\cite{Angulo:2015eqa} for more details. In this section and for the rest of the paper we will specialize in halos, so that we will substitute the subscript ${}_M$ for $_{h}$, because we have numerical data for halos. Our procedures equally apply to galaxies. The halo density field reads:
\bea\label{eq:euler_bias_3}
&& \delta_h(\vec x,t) \simeq \int^t dt' \; H(t') \; \left[   \bar c_{\delta}(t,t')\; :\delta(\xfl, t'): \right. 
\\
&& \qquad + \bar c_{\delta^2}(t,t')\;  :\delta(\xfl,t')^2:  +\bar c_{s^2}(t,t')\;  :s^2(\xfl, t') :\nonumber
\\
&& \qquad + \bar c_{\delta^3}(t,t') \;   :\delta(\xfl, t')^3 : + \bar c_{\delta s^2}(t,t') \;   : \delta(\xfl,t') s^2(\xfl,t'): + \bar c_{\psi}(t,t') \;   :\psi(\xfl,t'): \nonumber
\\
&& \qquad \qquad + \bar c_{st}(t,t') \;   :st(\xfl,t'): + \bar c_{s^3}(t,t') \;     :s^3(\xfl, t'): \nonumber
\\
&& \qquad + \bar c_{\epsilon}(t,t') \; \epsilon(\xfl, t') + \bar c_{\epsilon \delta}(t,t') \; :\epsilon(\xfl, t') \delta(\xfl, t'): \nonumber
\\
&& \qquad + \bar{c}_{\partial^2 \delta}(t, t'): \frac{\partial^2_{x_{\rm fl}}}{k^2_M} \delta(\bold{x}_{\rm fl}, t') :+ \bar{c}_{\partial^2 \delta^2}(t, t') :\frac{\partial^2_{x_{\rm fl}}}{k^2_M} \delta^2(\bold{x}_{\rm fl}, t'): + \bar{c}_{\partial^2 s^2}(t, t'): \frac{\partial^2_{x_{\rm fl}}}{k^2_M} s^2(\bold{x}_{\rm fl}, t'): \nonumber
\\
&& \qquad \qquad + \bar{c}_{(\partial \delta)^2}(t, t') :\frac{\partial_{x_{\rm fl},i}}{k_M} \delta(\bold{x}_{\rm fl}, t') \frac{\partial_{x_{\rm fl}}^i}{k_M} \delta(\bold{x}_{\rm fl}, t') : + \bar{c}_{\partial^4 \delta}(t, t') \frac{\partial^4_{x_{\rm fl}}}{k_M^4} \delta(\bold x_{\rm fl}, t') \nonumber
\\
&& \left. \quad + \bar c_{\partial^2 \epsilon}(t, t') \frac{\partial_{x_{\rm fl}}^2}{k_M^2} \epsilon(\xfl, t') + \bar c_{\partial^2 \epsilon \delta}(t,t') : \left( \frac{\partial_{x_{\rm fl}}^2}{k_M^2} \epsilon(\xfl, t') \right) \delta(\xfl, t') :+ \bar c_{\epsilon \partial^2 \delta}(t, t') :\epsilon(\xfl, t') \frac{\partial_{x_{\rm fl}}^2}{k_M^2} \delta(\xfl, t') : \right]. \nonumber
\eea
The notation is the same as in \cite{Angulo:2015eqa}, but let us remind it here. The field $\xfl$ is defined iteratively by
\beq\label{eq:xfl}
\xfl(\vec x, \tau,\tau') = \vec x - \int_{\tau'}^{\tau} d\tau''\; \vec{v}(\tau'',\vec x_{\rm fl}(\vec x, \tau,\tau''))\ .
\eeq
Then, the notation $:{\cal O}:$ means the normal ordering of operator $\cal O$:
\be
:{\cal O}: = \cal O  -\langle \cal O \rangle.
\ee
The field $\theta$ is defined by $\theta\equiv\d_i\tilde v^i=-\frac{D}{D'} \d_i v^i$. From the continuity equation at linear order, we have $\d_i v^i=-\frac{D'}{D} \delta$, where $'=\d/\d\tau$,
$\tau$ being the conformal time. Hence, at linear order, $\theta = \delta$. 
Furthermore, the gravitational field $\phi$ is redefined for convenience so that the Poisson equation becomes $\d^2\phi=\delta$.
Following~\cite{McDonald:2009dh}, we define new variables designed to vanish at low order, making their treatment simpler.
The first new independent variable is
\be
\eta(\vec x,t)=\theta(\vec x,t)- \delta(\vec x,t)\ ,
\ee
which vanish at linear order. Then, instead of using the gravitational field $\phi$, it is convenient to define the traceless tidal tensor 
\be
s_{ij}(\vec x,t)=\d_i\d_j\phi(\vec x,t)-\frac{1}{3} \delta_{ij}\, \delta(\vec x,t) \ .
\ee
Also vanishing at linear order is the new variable
\be
t_{ij}(\vec x,t)=\d_i \tilde{v}_j(\vec x,t)-\frac{1}{3}\delta_{ij} \theta(\vec x,t) -s_{ij}(\vec x,t)\ .
\ee
Finally the last variable is defined as 
\be
\psi(\vec x,t)=\eta(\vec x,t)-\frac{2}{7} s^2(\vec x,t)+\frac{4}{21}\delta(\vec x,t)^2\ ,
\ee
such that it contributes only from the third order in perturbation theory. In eq.~\eqref{eq:euler_bias_3}, we used the notations $s^2(\xfl,t)= s_{ij}(\xfl,t')s^{ij}(\xfl,t')$, $s^3(\xfl,t)=s_{ij}(\xfl,t')  s^{il}(\xfl,t')s_l{}^{j}(\xfl,t')$ and $st(\xfl,t)=s_{ij}(\xfl,t')  t^{ij}(\xfl,t')$, where indices are lowered and raised with $\delta^{ij}$. 

The last three lines of \eqref{eq:euler_bias_3} represent the contribution from the higher derivative terms. The scale~$\km$ here is the comoving wavenumber enclosing the mass of an object~\cite{Senatore:2014eva}. We approximatively have $k_M \sim 2 \pi \left( \frac{4 \pi}{3} \frac{\rho_b,0}{M} \right)^{1/3}$. This estimate implies that the terms proportional to $k^2/k_M^2$ are more important for objects with large mass. Only the first term was present in \cite{Angulo:2015eqa}, but, as we argued in the introduction, more of them are needed to predict the clustering of high mass halos as accurately as the clustering of smaller mass halos.
We should also consider the terms $\d_i \d_j s^{ij}$, $\d_i \d_j \d_k \phi \, \d^i \d^j \d^k \phi$, $\d_i \d_j v_k \, \d^j \d^j v^k$, $\d_i \d_j v_k \, \d^i \d^k v^j$, $\d_i \d_j t^{ij}$ and $\d_i \epsilon \, \d^i \delta$,  but, at the order at which we work, their contribution is degenerate with the higher-derivative terms that we explicitly write in (\ref{eq:euler_bias_3}), so we do not write them to keep the expression as light as possible (for more details see below and eq.~\eqref{eq:higher_deri} in App.~\ref{app:operators}, where the determination of a basis of linearly independent contributions is done following~\cite{Angulo:2015eqa}). Also, we omitted the stochastic operators $\epsilon s$, $\epsilon t$, etc. since, as argued in \cite{Angulo:2015eqa}, they have a negligible impact at the order at which we are working in this paper. Also be aware that, as discussed in the introduction, at the order at which we work, not every higher derivative term will be used in every observables: the two power-spectra will contain only $\frac{\d^2}{k_M^2} \delta, \frac{\d^4}{k_M^4} \delta$ and $\frac{\d^2}{k_M^2} \epsilon$, whereas the three bispectra, which are computed at tree level, will contain every term except $\frac{\d^4}{k_M^4} \delta$.

As in~\cite{Senatore:2014eva,Angulo:2015eqa}, we expand each field perturbatively and use our knowledge of time dependence to symbolically compute the time integral in \eqref{eq:euler_bias_3}, which gives:
\bea\label{eq:euler_bias_4}
&& \delta_h(\bold{k}, t) = \nonumber
\\
&& \quad = c_{\delta, 1}(t) \delta^{(1)}(\bold{k}, t) + c_{\delta, 2}(t) \delta^{(2)}(\bold{k}, t) + c_{\delta, 3}(t) \delta^{(3)}(\bold{k}, t) + c_{\delta, 3_{c_s}}(t) \delta^{(3)}_{c_s}(\bold{k}, t) \nonumber
\\
&& \qquad + [c_{\delta, 1}(t) - c_{\delta, 2}(t)] \left[\partial_i \delta^{(1)} \frac{\partial^i}{\partial^2} \theta^{(1)} \right]_{\bold{k}}(t) + [c_{\delta, 2}(t) - c_{\delta, 3}(t)] \left[\partial_i \delta^{(2)} \frac{\partial^i}{\partial^2} \theta^{(1)} \right]_{\bold{k}}(t) \nonumber
\\
&& \qquad + \frac{1}{2} [c_{\delta, 1}(t) - c_{\delta, 3}(t)] \left[\partial_i \delta^{(1)} \frac{\partial^i}{\partial^2} \theta^{(2)} \right]_{\bold{k}}(t) \nonumber
\\
&& \qquad + \frac{1}{2} [c_{\delta, 1}(t) - 2c_{\delta, 2}(t) + c_{\delta, 3}(t)] \nonumber
\\
&& \qquad \times \left( \left[\partial_i \delta^{(1)} \frac{\partial_j \partial^i}{\partial^2} \theta^{(1)} \frac{\partial^j}{\partial^2} \theta^{(1)} \right]_{\bold{k}}(t) + \left[\partial_i \partial_j \delta^{(1)} \frac{\partial^i}{\partial^2} \theta^{(1)} \frac{\partial^j}{\partial^2} \theta^{(1)} \right]_{\bold{k}}(t) \right) \nonumber
\\
&& \qquad + c_{\delta^2, 1}(t) [\delta^2]^{(2)}_{\bold{k}}(t) + c_{\delta^2, 2}(t) [\delta^2]^{(3)}_{\bold{k}}(t) + 2[c_{\delta^2, 1}(t) - c_{\delta^2, 2}(t)] \left[ \delta^{(1)} \partial_i \delta^{(1)} \frac{\partial^i}{\partial^2} \theta^{(1)} \right]_{\bold{k}}(t) \nonumber
\\
&& \qquad + c_{s^2, 1}(t) [s^2]^{(2)}_{\bold{k}}(t) + c_{s^2, 2}(t) [s^2]^{(3)}_{\bold{k}}(t) + 2[c_{s^2, 1}(t) - c_{s^2, 2}(t)] \left[ s_{lm}^{(1)} \partial_i (s^{lm})^{(1)} \frac{\partial^i}{\partial^2} \theta^{(1)} \right]_{\bold{k}}(t) \nonumber
\\
&& \qquad + c_{\delta^3}(t) [\delta^3]^{(3)}_{\bold k}(t) + c_{s^3}(t) [s^3]^{(3)}_{\bold{k}}(t) + c_{st}(t) [st]^{(3)}_{\bold{k}}(t) + c_{\psi}(t) \psi^{(3)}(\bold k, t) + c_{\delta s^2}(t) [\delta s^2]^{(3)}_{\bold{k}}(t) \nonumber
\\
&& \qquad  + [\epsilon]_{\bold k}(t) + c_{\epsilon \delta}(t) [\epsilon \delta]^{(1)}_{\bold k}(t) \nonumber
\\
&& \qquad - c_{\partial^2 \delta, 1}(t) \frac{k^2}{k_M^2} \delta^{(1)}(\bold{k}, t) + [c_{\partial^2 \delta, 1}(t) - c_{\partial^2 \delta, 2}(t))] \left[ \left(\frac{\partial^2}{k_M^2} \partial_i \delta^{(1)} \right) \frac{\partial^i}{\partial^2} \theta^{(1)} \right]_{\bold{k}}(t) \nonumber
\\
&& \qquad - c_{\partial^2 \delta, 2}(t) \frac{k^2}{k_M^2} \delta^{(2)}(\bold{k}, t) - c_{\partial^2 \delta^2}(t) \frac{k^2}{k_M^2} [\delta^2]^{(2)}_{\bold{k}}(t) - c_{\partial^2 s^2}(t) \frac{k^2}{k_M^2} [s^2]^{(2)}_{\bold{k}}(t) \nonumber
\\
&& \qquad + c_{(\partial \delta)^2}(t) \left[ \frac{\partial_i}{k_M} \delta^{(1)} \frac{\partial^i}{k_M} \delta^{(1)} \right]_{\bold{k}}(t) + c_{\partial^4 \delta}(t) \frac{k^4}{k_M^4} \delta^{(1)}(\bold k, t) \nonumber
\\
&& \qquad - c_{\d^2 \epsilon}(t) \frac{k^2}{k_M^2} [\epsilon]_{\bold k}(t) + c_{\d^2 \epsilon \delta}(t) \left[ \left( \frac{\d^2}{k_M^2} \epsilon \right) \delta \right]^{(1)}_{\bold k}(t) + c_{\epsilon \d^2 \delta}(t) \left[ \epsilon \frac{\d^2}{k_M^2} \delta \right]^{(1)}_{\bold k}(t).
\eea
In front of terms of second order, $\d^2_{x_{\rm fl}}$ has been replaced by $\d^2_x$, because, at the order at which we work, their contribution would give rise to third order terms, which, to the order at which we work, contribute only to the power spectrum at order $\epsilon_M \epsilon_{\delta}^2$, which is negligible for us. Furthermore, notice that at lines 7 and 8, the sign $+2$ is not the same as in \cite{Angulo:2015eqa}. The correction of this typo changes some numerical values of the bias coefficient throughout the paper but it does not affect any of the results, as, at the order at which we work, its contribution is degenerate with their numerical value.

This expression is more easily understood when separated into a sum of bias coefficients multiplying there respective operators, split into different perturbative orders labelled by $^{(n)}$,
\bea \label{eq:CoI_all}
\delta_h(\bold k, t) & = &  c_{\delta, 1}(t) \left[ \mathbb{C}^{(1)}_{\delta, 1}(\bold k, t) + \mathbb{C}^{(2)}_{\delta, 1}(\bold k, t) + \mathbb{C}^{(3)}_{\delta, 1}(\bold k, t) \right] \nonumber
\\
& + & c_{\delta, 2}(t) \left[ \qquad \qquad \quad \, \mathbb{C}^{(2)}_{\delta, 2}(\bold k, t) + \mathbb{C}^{(3)}_{\delta, 2}(\bold k, t) \right] \nonumber
\\
& + & c_{\delta, 3}(t) \left[\qquad \qquad \qquad \qquad \qquad \, \, \mathbb{C}^{(3)}_{\delta, 3}(\bold k, t) \right] \nonumber
\\
& + & c_{\delta, 3_{c_s}}(t) \, \mathbb{C}^{(3)}_{\delta, 3_{c_s}}(\bold k, t) \nonumber
\\
& + & c_{\delta^2, 1}(t) \left[ \mathbb{C}^{(2)}_{\delta^2, 1}(\bold k, t) + \mathbb{C}^{(3)}_{\delta^2, 1}(\bold k, t) \right] + c_{\delta^2, 2}(t) \, \mathbb{C}^{(3)}_{\delta^2, 2}(\bold k, t) \nonumber
\\
& + & c_{s^2, 1}(t) \left[ \mathbb{C}^{(2)}_{s^2, 1}(\bold k, t) + \mathbb{C}^{(3)}_{s^2, 1}(\bold k, t) \right] + c_{s^2, 2}(t) \, \mathbb{C}^{(3)}_{s^2, 2}(\bold k, t) \nonumber
\\
& + & c_{\delta^3}(t) \, \mathbb{C}^{(3)}_{\delta^3}(\bold k, t) + c_{s^3}(t) \, \mathbb{C}^{(3)}_{s^3}(\bold k, t) + c_{st}(t) \, \mathbb{C}^{(3)}_{st}(\bold k, t) + c_{\psi}(t) \, \mathbb{C}^{(3)}_{\psi}(\bold k, t) + c_{\delta s^2}(t) \, \mathbb{C}^{(3)}_{\delta s^2}(\bold k, t) \nonumber
\\
& + & \mathbb{C}_{\epsilon}(\bold k, t) + c_{\epsilon \delta}(t) \, \mathbb{C}^{(1)}_{\epsilon \delta}(\bold k, t) \nonumber
\\
& + & c_{\partial^2 \delta, 1}(t) \left[ \mathbb{C}^{(1)}_{\partial^2 \delta, 1}(\bold k, t) + \mathbb{C}^{(2)}_{\partial^2 \delta, 1}(\bold k, t) \right] + c_{\partial^2 \delta, 2}(t) \, \mathbb{C}^{(2)}_{\partial^2 \delta, 2}(\bold k, t) \nonumber
\\
& + & c_{\partial^2 \delta^2}(t) \, \mathbb{C}^{(2)}_{\partial^2 \delta^2}(\bold k, t) + c_{\partial^2 s^2}(t) \, \mathbb{C}^{(2)}_{\partial^2 s^2}(\bold k, t) + c_{(\partial \delta)^2}(t) \, \mathbb{C}^{(2)}_{(\partial \delta)^2}(\bold k, t) + c_{\partial^4 \delta}(t) \, \mathbb{C}^{(1)}_{\partial^4 \delta}(\bold k, t) \nonumber
\\
& + & c_{\d^2 \epsilon}(t) \, \mathbb{C}_{\d^2 \epsilon}(\bold k, t) + c_{\d^2 \epsilon \delta}(t) \, \mathbb{C}^{(1)}_{\d^2 \epsilon \delta}(\bold k, t) + c_{\epsilon \d^2 \delta}(t) \, \mathbb{C}^{(1)}_{\epsilon \d^2 \delta}(\bold k, t),
\eea 
Explicit expressions for the operators $\mathbb{C}$ appearing in eq.~\eqref{eq:CoI_all} are provided in App.~\ref{app:operators}, eq.~(\ref{eq:C_all_explicit}).

As in \cite{Angulo:2015eqa}, we will now get rid of the many redundancies of eq.~\eqref{eq:CoI_all}. Indeed, several of the operators are degenerate with one another, and it is possible to construct a basis of linearly independent operators. We will choose the basis following the same procedure as in~\cite{Angulo:2015eqa}, giving rise to the so-called Basis of Descendants ({\it BoD} basis), where the bias operators that appear for the first time in the expansion at a given 
perturbative order are kept only if they are not degenerate with any of the bias operators introduced earlier. This yields
\bea \label{eq:CoI}
\rm 1st\,order : &&\left\{ \mathbb{C}^{(1)}_{\delta, 1} \right\}, \nonumber
\\
\rm 2nd\,order : &&\left\{ \mathbb{C}^{(2)}_{\delta, 1}, \mathbb{C}^{(2)}_{\delta, 2}, \mathbb{C}^{(2)}_{\delta^2, 1} \right\}, \nonumber
\\
\rm 3rd\,order : &&\left\{ \mathbb{C}^{(3)}_{\delta, 1}, \mathbb{C}^{(3)}_{\delta, 2}, \mathbb{C}^{(3)}_{\delta, 3}, \mathbb{C}^{(3)}_{\delta^2, 1}, \mathbb{C}^{(3)}_{\delta^2, 2}, \mathbb{C}^{(3)}_{s^2, 2}, \mathbb{C}^{(3)}_{\delta^3}, \mathbb{C}^{(3)}_{\delta, 3_{c_s}} \right\}, \nonumber
\\
\rm 1st\,order\,higher\,derivative : &&\left\{{\mathbb{C}^{(1)}_{\partial^2 \delta, 1},} \mathbb{C}^{(1)}_{\partial^4 \delta} \right\}, \nonumber
\\
\rm 2nd\,order\,higher\,derivative : &&\left\{ \mathbb{C}^{(2)}_{\partial^2 \delta, 1},  \mathbb{C}^{(2)}_{\partial^2 \delta, 2}, \mathbb{C}^{(2)}_{\partial^2 \delta^2}, \mathbb{C}^{(2)}_{\partial^2 s^2}, \mathbb{C}^{(2)}_{(\partial \delta)^2} \right\} \nonumber
\\
\text{Stochastic: }&&\left\lbrace\mathbb{C}_{\epsilon},~\mathbb{C}^{(1)}_{\df\epsilon,1}\right\rbrace\ \nonumber
\\
\text{Higher order stochastic :} &&\left\{ \mathbb{C}_{\d^2 \epsilon}, \mathbb{C}^{(1)}_{\d^2 \epsilon \delta}, \mathbb{C}^{(1)}_{\epsilon \d^2 \delta} \right\}.
\eea
The linear combinations relating each degenerate operator with the ones in the basis are shown in App.~\ref{app:operators}, eq.~(\ref{eq:C_all_explicit_4}). We refer to \cite{Angulo:2015eqa} for the procedure to compute the degeneracies. Note that even though $\mathbb{C}^{(1)}_{\d^2 \delta, 1}$ is degenerate with $\mathbb{C}^{(3)}_{\delta, 3_{c_s}}$, it will appear in the bispectra, because they do not contain~$\mathbb{C}^{(3)}_{\delta, 3_{c_s}}$. The fact that $\mathbb{C}^{(1)}_{\partial^2 \delta, 1}$ appears in the bispectra justifies that we include it in equation~\eqref{eq:CoI}, despite its degeneracy with $\mathbb{C}^{(3)}_{\delta, 3_{c_s}}$.
  
After this simplification the overdensity field up to third order in perturbation theory becomes
\bea \label{eq:delta_h_CoI_t}
\delta_h(\vec k,t) & = & \tilde{c}_{\delta,1}(t) \; \Big( \mathbb{C}^{(1)}_{\delta, 1}(\vec k, t )+ \mathbb{C}^{(2)}_{\delta, 1}(\vec k, t) + \mathbb{C}^{(3)}_{\delta, 1}(\vec k, t) \Big) \nonumber
\\
& + & \tilde{c}_{\delta, 2}(t) \; \Big( \mathbb{C}^{(2)}_{\delta, 2}(\vec k, t) + \mathbb{C}^{(3)}_{\delta, 2}(\vec k, t) \Big) \nonumber
\\
& + & \tilde{c}_{\delta, 3}(t) \; \mathbb{C}^{(3)}_{\delta, 3}(\vec k, t)
+ \tilde{c}_{\delta, 3_{c_s}}(t) \; \mathbb{C}^{(3)}_{\delta, 3_{c_s}}(\vec k, t) \nonumber
\\
& + & \tilde{c}_{\delta^2, 1}(t) \; \Big( \mathbb{C}^{(2)}_{\delta^2, 1}(\vec k, t) + \mathbb{C}^{(3)}_{\delta^2, 1}(\vec k, t) \Big) \nonumber
\\
& + & \tilde{c}_{\delta^2, 2}(t) \; \mathbb{C}^{(3)}_{\delta^2 ,2}(\vec k, t) + \tilde{c}_{s^2, 2}(t) \; \mathbb{C}^{(3)}_{s^2, 2}(\vec k, t) \nonumber
\\
& + & \tilde{c}_{\delta^3}(t) \; \mathbb{C}^{(3)}_{\delta^3}(\vec k, t) \nonumber
\\
& + & \mathbb{C}_{\epsilon}(\vec k,t) + \tilde{c}_{\epsilon \delta}(t) \mathbb{C}^{(1)}_{\epsilon \delta}(\bold k, t) \nonumber
\\
& + & \tilde{c}_{\partial^2 \delta, 1}(t) \; \mathbb{C}^{(2)}_{\partial^2 \delta, 1}(\bold k, t) + \tilde{c}_{\partial^2 \delta, 2}(t) \, \mathbb{C}^{(2)}_{\partial^2 \delta, 2}(\bold k, t) \nonumber
\\
& + & \tilde{c}_{\partial^2 \delta^2}(t) \, \mathbb{C}^{(2)}_{\partial^2 \delta^2}(\bold k, t) + \tilde{c}_{\partial^2 s^2}(t) \, \mathbb{C}^{(2)}_{\partial^2 s^2}(\bold k, t) + \tilde{c}_{(\partial \delta)^2}(t) \, \mathbb{C}^{(2)}_{(\partial \delta)^2}(\bold k, t) + \tilde{c}_{\partial^4 \delta}(t) \, \mathbb{C}^{(1)}_{\partial^4 \delta}(\bold k, t) \nonumber
\\
& + & \tilde{c}_{\d^2 \epsilon}(t) \, \mathbb{C}_{\d^2 \epsilon}(\bold k, t) + \tilde{c}_{\d^2 \epsilon \delta}(t) \, \mathbb{C}^{(1)}_{\d^2 \epsilon \delta}(\bold k, t) + \tilde{c}_{\epsilon \d^2 \delta}(t) \, \mathbb{C}^{(1)}_{\epsilon \d^2 \delta}(\bold k, t),
\eea
where the new bias coefficients $\tilde{c}_i$ are linear combinations of the old ones, see eq.~(\ref{eq:c_tildas}) of App.~\ref{app:operators} for the explicit expressions.

In order to further simplify the computations, it is useful to organize the perturbative expansion similarly to what we ordinarily do for the dark matter overdensity field, where we write:
\bea \label{eq:CoI_all_dm}
\delta(\vec k,t)&=& \delta^{(1)} (\vec k,t) + \delta^{(2)} (\vec k,t) + \delta^{(3)} (\vec k,t) + \delta^{(3)}_{c_s} (\vec k,t) + \ldots\ ,
\eea
where we have 
\begin{align} \label{eq:CoI_all_kern}
\delta^{(1)} (\vec k,t)&= \int \frac{d^3q_1}{(2\pi)^3}\;F^{(1)}_s (\vec q_1)\,\delta_D^{(3)}(\vec k- \vec q_1)\,\delta^{(1)} (\vec q_1,t)\ , \\ \nonumber
\delta^{(2)} (\vec k,t)&= \int \frac{d^3q_1}{(2\pi)^3}\;\frac{d^3q_2}{(2\pi)^3}\;\; 
F^{(2)}_s (\vec q_1, \vec q_2)\,\delta_D^{(3)}(\vec k-\vec q_1-\vec q_2)\,\delta^{(1)} (\vec q_1,t)\delta^{(1)} (\vec q_2,t)\ , \\ \nonumber
\delta^{(3)} (\vec k,t)&= \int \frac{d^3q_1}{(2\pi)^3}\;\frac{d^3q_2}{(2\pi)^3}\;\frac{d^3q_3}{(2\pi)^3}\;\; 
F^{(3)}_s (\vec q_1,\vec q_2,\vec q_3)\,\delta_D^{(3)}(\vec k-\vec q_1-\vec q_2-\vec q_3)\,\delta^{(1)} (\vec q_1,t)\delta^{(1)} (\vec q_2,t)\delta^{(1)} (\vec q_3,t)\ ,
\end{align}
where the $F^{(n)}_s$ functions are standard Eulerian perturbation theory kernels (see e.g.~\cite{Bernardeau:2001qr} for definitions).
By analogy, for the overdensity field of biased tracers it is useful to rewrite (\ref{eq:CoI_all}) as 
\bea \label{eq:CoI_all_hl}
\delta_h(\vec k,t)&=& \delta^{(1)}_h (\vec k,t) + \delta^{(2)}_h (\vec k,t) + \delta^{(3)}_h (\vec k,t) + \delta^{(3)}_{h, c_s} (\vec k,t) + \ldots + (\epsilon\text{-terms})\ ,
\eea
where we have introduced the new kernels for biased tracers 
\begin{align} \label{eq:CoI_all_hl_kern}
\delta^{(1)}_h (\vec k,t)&= \int \frac{d^3q_1}{(2\pi)^3}\;K^{(1)}_s (\vec q_1)\,\delta_D^{(3)}(\vec k-\vec q_1)\, \delta^{(1)} (\vec q_1,t)\ , \\ \nonumber
\delta^{(2)}_h (\vec k,t)&= \int \frac{d^3q_1}{(2\pi)^3}\;\frac{d^3q_2}{(2\pi)^3}\;\; 
K^{(2)}_s (\vec q_1,\vec q_2)\,\delta_D^{(3)}(\vec k-\vec q_1-\vec q_2)\,\delta^{(1)} (\vec q_1,t)\delta^{(1)} (\vec q_2,t)\ , \\ \nonumber
\delta^{(3)}_h (\vec k,t)&= \int \frac{d^3q_1}{(2\pi)^3}\;\frac{d^3q_2}{(2\pi)^3}\;\frac{d^3q_3}{(2\pi)^3}\;\; 
K^{(3)}_s (\vec q_1,\vec q_2,\vec q_3)\,\delta_D^{(3)}(\vec k-\vec q_1-\vec q_2-\vec q_3)\,\delta^{(1)} (\vec q_1,t)\delta^{(1)} (\vec q_2,t)\delta^{(1)} (\vec q_3,t)\ .
\end{align}
The kernels $K^{(i)}_s$ can be expressed in the \textit{BoD} basis as 
\bea \label{eq:CoI}
K^{(1)}_s (\vec q_1, t) & = & \tilde{c}_{\delta, 1}(t) \; \widehat{c}^{(1)}_{\delta, 1}(\vec q_1)+ { \tilde{c}_{\d^2 \delta,1}(t) \,  \widehat{c}^{(1)}_{\d^2 \delta,1}(\vec q_1)} + \tilde{c}_{\d^4 \delta}(t) \, \widehat{c}^{(1)}_{\d^4 \delta}(\vec q_1),  \nonumber
\\
K^{(2)}_s (\vec q_1, \vec q_2, t) & = & \tilde{c}_{\delta, 1}(t) \; \widehat{c}^{(2)}_{{\delta, 1}}(\vec q_1, \vec q_2) + \tilde{c}_{\delta, 2}(t) \; \widehat{c}^{(2)}_{{\delta,2}}(\vec q_1, \vec q_2) + \tilde{c}_{\delta^2, 1}(t) \; \widehat{c}^{(2)}_{{\delta^2,1}}(\vec q_1, \vec q_2) \nonumber
\\
&& + \tilde{c}_{\d^2 \delta, 1}(t) \; \widehat{c}^{(2)}_{{\d^2 \delta, 1}}(\vec q_1, \vec q_2 + \tilde{c}_{\d^2 \delta, 2}(t) \; \widehat{c}^{(2)}_{{\d^2 \delta, 2}}(\vec q_1, \vec q_2)) + \tilde{c}_{\d^2 \delta^2}(t) \; \widehat{c}^{(2)}_{{\d^2 \delta^2}}(\vec q_1, \vec q_2) \nonumber
\\
&& + \tilde{c}_{\d^2 s^2}(t) \; \widehat{c}^{(2)}_{{\d^2 s^2}}(\vec q_1, \vec q_2) + \tilde{c}_{(\d \delta)^2}(t) \; \widehat{c}^{(2)}_{{(\d \delta)^2}}(\vec q_1, \vec q_2), \nonumber
\\
K^{(3)}_s (\vec q_1, \vec q_2, \vec q_3, t) & = & \tilde{c}_{\delta, 1}(t) \; \widehat{c}^{(3)}_{{\delta, 1}}(\vec q_1, \vec q_2, \vec q_3) + \tilde{c}_{\delta, 2}(t) \; \widehat{c}^{(3)}_{{\delta, 2}}(\vec q_1, \vec q_2, \vec q_3) + \tilde{c}_{\delta, 3}(t) \; \widehat{c}^{(3)}_{{\delta, 3}}(\vec q_1, \vec q_2, \vec q_3) \nonumber
\\
&& + \tilde{c}_{\delta^2, 1}(t) \; \widehat{c}^{(3)}_{{\delta^2, 1}}(\vec q_1,\vec q_2 ,\vec q_3) + \tilde{c}_{\delta^2, 2}(t) \; \widehat{c}^{(3)}_{{\delta^2, 2}}(\vec q_1, \vec q_2, \vec q_3)  \nonumber
\\
&& + \tilde{c}_{\delta^3}(t) \; \widehat{c}^{(3)}_{{\delta^3}}(\vec q_1, \vec q_2, \vec q_3) + \tilde{c}_{s^2, 2}(t) \; \widehat{c}^{(3)}_{{s^2, 2}}(\vec q_1,\vec q_2, \vec q_3).
\eea
We notice that the kernels $K^{(i)}_s$ are linear combinations of bias parameters $\tilde c_{\dots}(t)$, multiplied by subkernels $\hat c^{(n)}_{\dots}(q_1,q_2,\ldots)$, 
that are defined in eq.~(\ref{eq:c_all_explicit}) of Appendix \ref{app:operators}.

As we discussed in the introduction, the fact that the power spectra and the bispectra are computed to different order implies that not all the higher derivative biases appear in all observables we consider. In particular, in $K_s^{(2)}(\bold q_1, \bold q_2, t)$ the higher derivative terms must appear only in the bispectra and not the power spectra, since in the power spectrum we only need to add a higher derivative linear terms. Similarly, the higher derivative term in $K_s^{(1)}(\bold q_1, t)$ will appear only in the power spectra and not in the bispectra.

\subsection{One-loop power spectra and tree-level bispectra}
\label{subsec:obs}

Having written the expansion of the tracer overdensity fields, we are ready to compute its correlation functions. The observables we are going to compute are the halo-matter 
cross power spectrum $P_{hm}$ and the halo-halo auto power spectrum, $P_{hh}$, at one-loop order, 
as well as the halo-matter-matter bispectrum $B_{hmm}$, the  halo-halo-matter bispectrum, $B_{hhm}$,
and halo-halo-halo bispectrum, $B_{hhh}$, at tree level.

The halo-matter cross power spectrum $P_{hm}$ at one-loop order is
\bea \label{eq:Phm}
P_{hm}(k, t) & = & \langle \delta_h^{(1)}(\bold k, t) \delta^{(1)}(\bold k, t) \rangle' + \langle \delta_h^{(2)}(\bold k, t) \delta^{(2)}(\bold k, t) \rangle' + \langle \delta_h^{(3)}(\bold k, t) \delta^{(1)}(\bold k, t) \rangle' \nonumber
\\
&& \qquad + \langle \delta_h^{(1)}(\bold k, t) \delta^{(3)}(\bold k, t) \rangle' + \langle \delta_{h,c_s}^{(3)}(\bold k, t) \delta^{(1)}(\bold k, t) \rangle' + \langle \delta_h^{(1)}(\bold k, t) \delta^{(3)}_{c_s}(\bold k, t) \rangle' \nonumber
\\
& = & \left( \tilde{c}_{\delta, 1}(t) + \frac{k^4}{k_M^4} \tilde{c}_{\partial^4 \delta}(t) \right) P_{11}(k;t,t) \nonumber
\\
&& + 2\int \frac{d^3 q}{(2 \pi)^3} K_s^{(2)}(\bold k - \bold q, \bold q) F_s^{(2)}(\bold k - \bold q, \bold q) P_{11}(q;t,t) P_{11}(|\bold k - \bold q|;t,t) \nonumber
\\
&& + 3 P_{11}(k;t,t) \int \frac{d^3 q}{(2 \pi)^3} \left(K_s^{(3)}(\bold k, - \bold q, \bold q) + \tilde{c}_{\delta, 1}(t) F_s^{(3)}(\bold k, - \bold q, \bold q) \right) P_{11}(q;t,t) \nonumber
\\
&& + \left(\tilde{c}_{\delta, 3_{c_s}}(t) + \tilde{c}_{\delta, 1}(t) \right) \left( - (2 \pi) c^2_{s(1)}(t) \right) \frac{k^2}{k_{NL}^2} P_{11}(k;t,t)
\eea
where the primed brackets $\langle \rangle^{\prime}$ means that we have dropped the mumentum conserving Dirac $\delta$-function and the $(2 \pi)^3$ factor from the expectation value. Notice that since the kernels $K^{(n)}_s$ contain unknown bias parameters, they need to be 
expanded in order to be evaluated numerically. We will do this later on, after the renormalization. As mentioned earlier, when we will expand eq.~\eqref{eq:Phm}, we must discard the higher derivative term in $K_s^{(2)}(\bold q_1, \bold q_2, t)$.

Similarly, for the halo-halo auto power spectrum $P_{hh}$ we have:
\begin{eqnarray}\label{eq:Phh}
P_{hh}(k, t) & = & \langle \delta_h^{(1)}(\bold k, t) \delta_h^{(1)}(\bold k, t) \rangle' + \langle \delta_h^{(2)}(\bold k, t) \delta_h^{(2)}(\bold k, t) \rangle' + 2 \langle \delta_h^{(3)}(\bold k, t) \delta_h^{(1)}(\bold k, t) \rangle' \nonumber
\\
&& + \, 2 \langle \delta_h^{(1)}(\bold k, t) \delta_{h, c_s}^{(3)}(\bold k, t) \rangle' + \langle [\epsilon]_{\bold k}^{(1)} [\epsilon]_{\bold k}^{(1)} \rangle' + \tilde{c}_{\epsilon \delta}^2(t) \langle [\epsilon \delta]_{\bold k}^{(1)} [\epsilon \delta]_{\bold k}^{(1)} \rangle'    \nonumber
\\
&& + 2 \tilde{c}_{\partial^2 \epsilon}(t) \langle [\epsilon]_{\bold k} \left[ \frac{\partial^2}{k_M^2} \epsilon \right]_{\bold k} \rangle' \nonumber
\\
& = & \tilde{c}^2_{\delta, 1}(t) P_{11}(k) + 2 \tilde{c}_{\delta, 1}(t) \tilde{c}_{\partial^4 \delta}(t) \frac{k^4}{k_M^4} P_{11}(k) \nonumber
\\
&& + 2 \int \frac{d^3 q}{(2 \pi)^3} [K_s^{(2)}(\bold k - \bold q, \bold q)]^2 P_{11}(q) P_{11}(|\bold k - \bold q|) \nonumber
\\
&& + 6 \tilde{c}_{\delta, 1}(t) P_{11}(k) \int \frac{d^3 q}{(2 \pi)^3} K_s^{(3)}(\bold k, - \bold q, \bold q) P_{11}(q) \nonumber
\\
&& + 2 \tilde{c}_{\delta, 1}(t) \tilde{c}_{\delta, 3_{c_s}}(t) (-(2 \pi)) c_{s(1)}^2(t) \frac{k^2}{k_{NL}^2} P_{11}(k) + {\rm Const}_{\epsilon,P} \nonumber
\\
&& + \tilde{c}^2_{\epsilon \delta}(t) {\rm Const}_{\epsilon,P} \int \frac{d^3 q}{(2 \pi)^3} P_{11}(q) - 2 \frac{k^2}{k_M^2} \tilde{c}_{\partial^2 \epsilon}(t) {\rm Const}_{\epsilon,P}
\end{eqnarray}
We could have included the correlations $\langle [\epsilon \delta]^{(1)}_{\bold k} [\epsilon \frac{\partial^2}{k_M^2} \delta]^{(1)}_{\bold k} \rangle'$ and  $\langle [\epsilon \delta]^{(1)}_{\bold k} [\frac{\partial^2}{k_M^2} \epsilon \delta]^{(1)}_{\bold k} \rangle'$, but those contributions would then be absorbed by ${\rm Const}_{\epsilon,P}$ during the renormalization procedure, just like $\langle [\epsilon \delta]_{\bold k}^{(1)} [\epsilon \delta]_{\bold k}^{(1)} \rangle'$ does (see~\cite{Angulo:2015eqa}).

The bispectra are simply given by the correlation of three matter or halo fields. The three combinations we are interested in are:
\begin{eqnarray} \label{eq:Bhmm}
B_{hmm}(\bold k_1, \bold k_2, \bold k_3, t) & = & \langle \delta_h^{(2)}(\bold k_1, t) \delta^{(1)}(\bold k_2, t) \delta^{(1)}(\bold k_3, t) \rangle ' + \langle \delta_h^{(1)}(\bold k_1, t) \delta^{(2)}(\bold k_2, t) \delta^{(1)}(\bold k_3, t) \rangle '
\\
&& + \langle \delta_h^{(1)}(\bold k_1, t) \delta^{(1)}(\bold k_2, t) \delta^{(2)}(\bold k_3, t) \rangle ' \nonumber
\\
& = & 2 K_s^{(2)}(\bold k_2, \bold k_3, t) P_{11}(k_2;t,t) P_{11}(k_3;t,t) \nonumber
\\
&& + 2 K_s^{(1)}(\bold k_1, t) \left[ F_s^{(2)}(\bold k_1, \bold k_2) P_{11}(k_1;t,t) P_{11}(k_2;t,t) + F_s^{(2)}(\bold k_1, \bold k_3) P_{11}(k_1) P_{11}(k_3) \right] \ , \nonumber
\end{eqnarray}

\begin{eqnarray} \label{eq:Bhhm}
B_{hhm}(\bold k_1, \bold k_2, \bold k_3, t) & = & \langle \delta_h^{(2)}(\bold k_1, t) \delta_h^{(1)}(\bold k_2, t) \delta^{(1)}(\bold k_3, t) \rangle ' + \langle \delta_h^{(1)}(\bold k_1, t) \delta_h^{(2)}(\bold k_2, t) \delta^{(1)}(\bold k_3, t) \rangle ' \nonumber
\\
&& + \langle \delta_h^{(1)}(\bold k_1, t) \delta_h^{(1)}(\bold k_2, t) \delta^{(2)}(\bold k_3, t) \rangle' \nonumber
\\
&& + \tilde{c}_{\epsilon \delta}(t) \left( \langle [\epsilon]_{\bold k_1}^{(1)} [\epsilon \delta]^{(1)}_{\bold k_2} \delta^{(1)}(\bold k_3) \rangle' + \langle [\epsilon \delta]_{\bold k_1}^{(1)} [\epsilon]^{(1)}_{\bold k_2} \delta^{(1)}(\bold k_3) \rangle' \right) \nonumber
\\
&& + c_{\partial^2 \epsilon \delta}(t) \frac{1}{k_M^2} \left( \langle [\epsilon]^{(1)}_{\bold k_1} [\partial^2 \epsilon \delta]^{(1)}_{\bold k_2} \delta^{(1)}(\bold k_3) \rangle' + \langle [\partial^2 \epsilon \delta]_{\bold k_1}^{(1)} [\epsilon]^{(1)}_{\bold k_2} \delta^{(1)}(\bold k_3) \rangle'  \right) \nonumber
\\
&& + c_{\epsilon \partial^2 \delta}(t) \frac{1}{k_M^2} \left( \langle [\epsilon]^{(1)}_{\bold k_1} [\epsilon \partial^2 \delta]^{(1)}_{\bold k_2} \delta^{(1)}(\bold k_3) \rangle' + \langle [\epsilon \partial^2 \delta]_{\bold k_1}^{(1)} [\epsilon]^{(1)}_{\bold k_2} \delta^{(1)}(\bold k_3) \rangle' \right) \nonumber
\\
& = & 2 K_s^{(2)}(\bold k_2, \bold k_3, t) K_s^{(1)}(\bold k_2, t) P_{11}(k_2;t,t) P_{11}(k_3;t,t) \nonumber
\\
&& + 2 K_s^{(2)}(\bold k_1, \bold k_3, t) K_s^{(1)}(\bold k_1, t) P_{11}(k_1;t,t) P_{11}(k_3;t,t) \nonumber
\\
&& + 2 F_s^{(2)}(\bold k_1, \bold k_2) K_s^{(1)}(\bold k_1, t) K_s^{(1)}(\bold k_2, t) P_{11}(k_1;t,t) P_{11}(k_2;t,t) \nonumber
\\
&& + {\rm Const}_{\epsilon,P} P_{11}(k_3;t,t) \left( 2 \tilde{c}_{\epsilon \delta}(t) - \tilde{c}_{\partial^2 \epsilon \delta}(t) \frac{k_1^2 + k_2^2}{k_M^2} - 2 \tilde{c}_{\epsilon \partial^2 \delta}(t) \frac{k_3^2}{k_M^2} \right)\ ,
\end{eqnarray}

\begin{eqnarray} \label{eq:Bhhh}
B_{hhh}(\bold k_1, \bold k_2, \bold k_3, t) & = & \langle \delta_h^{(2)}(\bold k_1, t) \delta_h^{(1)}(\bold k_2, t) \delta_h^{(1)}(\bold k_3, t) \rangle '  + \langle [\epsilon]_{\bold k_1} [\epsilon]_{\bold k_2} [\epsilon]_{\bold k_3} \rangle' \\ \nn
&& +  \langle [\delta_h^{(1)}(\bold k_1, t)  [\epsilon]_{\bold k_2} [\epsilon\d^2\delta]_{\bold k_3} \rangle' +  \langle [\delta_h^{(1)}(\bold k_1, t) [\epsilon\d^2\delta]_{\bold k_2} [\epsilon]_{\bold k_3}  \rangle'  \nonumber
\\ \nn
&& +  \langle [\delta_h^{(1)}(\bold k_1, t)  [\epsilon]_{\bold k_2} [\d^2\epsilon\delta]_{\bold k_3} \rangle' +  \langle [\delta_h^{(1)}(\bold k_1, t) [\d^2\epsilon\delta]_{\bold k_2} [\epsilon]_{\bold k_3}  \rangle' + {\rm\; 2\ permutations}  \\ \nn
& = & \left\{ 2 K_s^{(2)}(\bold k_1, \bold k_2) K_s^{(1)}(\bold k_1) K_s^{(1)}(\bold k_2) P_{11}(k_1;t,t) P_{11}(k_2;t,t) + {\rm Const}_{\epsilon,P} P_{11}(k_1;t,t)  \right.\nonumber
\\ \nn
&& \left. \times \left( 2 \tilde{c}_{\delta,1}(t) \tilde{c}_{\epsilon \delta}(t) - 2 \frac{k_1^2}{k_M^2} \tilde{c}_{\partial^2 \delta, 1}(t) \tilde{c}_{\epsilon \delta}(t) - \frac{k_2^2 + k_3^2}{k_M^2} \tilde{c}_{\delta, 1}(t) \tilde{c}_{\partial^2 \epsilon \delta}(t) \right. \right. 
\\ \nn
&& \left. \left. - 2 \frac{k_1^2}{k_M^2} \tilde{c}_{\delta, 1}(t) \tilde{c}_{\epsilon \partial^2 \delta}(t) \right)  + 2 \, {\rm permutations}\right\} + {\rm Const}_{\epsilon,B} \  .
\end{eqnarray}
In the last two bispectra we omitted $\frac{\partial^2}{k_M^2} \epsilon$ because it gives a contribution degenerate with $\partial^2 \epsilon \delta$. Similarly, if we had included $\d_i \epsilon \, \d^i \delta$ in \eqref{eq:euler_bias_3}, its contribution would have been degenerated with $\epsilon \d^2 \delta$ after taking into account momentum conservation. In $B_{hhh}$ we do not include the contribution from $\langle [\d^2\epsilon]_{\bold k_1} [\epsilon]_{\bold k_2} [\epsilon]_{\bold k_3} \rangle'$, because, as we comment later, it does not appear to improve the fit to the data. In the case of the bispectrum, the kernel $K_s^{(2)}$ does contain every term appearing in eq.~\eqref{eq:CoI}, including the higher derivative ones.

\subsection{UV dependence and renormalization of bias coefficients}
\label{subsec:obs}

The momentum integrals in equations \eqref{eq:Phm} and \eqref{eq:Phh} span the full momentum-space even though perturbation theory is trusted only at low momenta. The bias parameters are able to re-absorb the error introduced by the high momentum contribution and to  faithfully implement the effect that short scales have on large scales. This procedure goes under the name of renormalization. Since the purpose of this paper is to show the importance of the higher derivative bias terms for the accuracy of the predictions for high mass tracers, which we include only at tree level, the renormalization procedure we need to perform is identical to the one done in~\cite{Angulo:2015eqa}, to which we refer for the details. The result is that the renormalization of the two power spectra is achieved by replacing the bias coefficients in the following way 
\begin{align}
\label{eq:renscheme}
&\tilde{c}_{\delta,1}\rightarrow b_{\delta,1}
- \sigma^2(t) \left(- \frac{13}{21} \tilde{c}_{\delta, 1} - \frac{34}{21} \tilde{c}_{\delta, 2} + 
 \frac{47}{21} \tilde{c}_{\delta, 3} - 2 \tilde{c}_{\delta^2, 1} + \frac{110}{21} \tilde{c}_{\delta^2, 2} + \frac{136}{63} \tilde{c}_{s^2, 2}
 + 3 \tilde{c}_{\delta^3}\right)
, \\ 
&\tilde{c}_{\delta,2}\rightarrow b_{\delta,2}\ ,\non\\ 
&\tilde{c}_{\delta,3}+15\tilde{c}_{s^2,2}\rightarrow b_{\delta,3}\ ,\non\\ 
&\tilde{c}_{\delta^2,1}\rightarrow b_{\delta^2}\ ,\non\\ 
&\tilde{c}_{\d^4 \delta}\rightarrow b_{\d^4 \delta}\ ,\non\\ 
&\tilde{c}_{\delta,3_{c_s}} \rightarrow b_{c_s}\ , \non\\
&\tilde{c}_{\d^2 \epsilon} \rightarrow b_{\d^2 \epsilon}\ ,\non\\ 
&{\rm \tilde{C}onst}_{\epsilon,P}\rightarrow {\rm Const}_{\epsilon}   \Big( 1 - \tilde{c}_{\epsilon \delta}^2 \sigma^2(t) \Big) \non\\ 
&\qquad\qquad\qquad   - 2\Sigma(t)^2 \left( \tilde{c}_{\delta,1}^2 + \tilde{c}_{\delta,2}^2 + \tilde{c}^2_{\delta^2,1} - 2 \; \tilde{c}_{\delta,1}\tilde{c}_{\delta,2}  
- 2 \; \tilde{c}_{\delta,1}\tilde{c}_{\delta^2,1} + 2 \; \tilde{c}_{\delta,2}\tilde{c}_{\delta^2,1} \right) 
\non\ .
\end{align}
Here every bias coefficient $b_i$ is intended to be a finite contribution. In \cite{Angulo:2015eqa} it was shown that the assumption that the stochastic bias is poisson distributed was sufficiently good to match the data. This assumptions relates the stochastic contribution ${\rm Const}_{\epsilon,B}$ to ${\rm Const}_{\epsilon}$ by the formula
\be 
{\rm Const}_{\epsilon,B} = {\rm Const}_{\epsilon}^2\ .
\ee
In terms of renormalized bias coefficients, the cross power spectra of halos and dark matter is now given by:
\begin{align}
 P_{hm}(k,t) =& b_{\delta,1}(t)\left( P_{11}(k;t,t) 
  +2\int \frac{d^3q}{(2\pi)^3} \;F^{(2)}_s(\vec{k}-\vec{q},\vec{q})\, \widehat{c}^{(2)}_{{\delta,1,s}}(\vec{k}-\q,\q)\, P_{11}(q;t,t)P_{11}( |\vec{k}-\vec{q}|;t,t) \right. \non\\
  &\qquad\qquad~~~~\left.+3 \; P_{11}(k;t,t) \int \frac{d^3q}{(2\pi)^3} \;\left(F^{(3)}_{s}(\vec{k},-\q,\q)+
  \widehat{c}^{(3)}_{{\delta,1,s}}(\vec{k},-\q,\q) + \sfrac{13}{63}\right)  P_{11}(q;t,t) \right) \non \\
  &+b_{\delta,2}(t) \; 2 \int \frac{d^3q}{(2\pi)^3} \;
                    F^{(2)}_s(\vec{k}-\vec{q},\vec{q})\left( F^{(2)}_s(\vec{k}-\vec{q},\vec{q}) - \widehat{c}^{(2)}_{{\delta,1,s}}(\vec{k}-\q,\q) \right)
                    P_{11}(q;t,t)P_{11}( |\vec{k}-\vec{q}|;t,t) \non \\
  &+b_{\delta, 3}(t)\; 3\; P_{11}(k;t,t) \int \frac{d^3q}{(2\pi)^3} \;\left( \widehat{c}^{(3)}_{{\delta,3,s}}(\vec{k},-\q,\q) - \sfrac{47}{63}\right)  P_{11}(q;t,t) \non \\
  &+b_{\delta^2}(t)\; 2 \int \frac{d^3q}{(2\pi)^3}F^{(2)}_s(\vec{k}-\vec{q},\vec{q})P_{11}(q;t,t)P_{11}( |\vec{k}-\vec{q}|;t,t)\non \\
  &+\left(b_{c_s}(t)+b_{\delta,1}(t) \right) (-2(2\pi))c^2_{s(1)}(t) \frac{k^2}{k^2_{\text{NL}}}P_{11}(k;t,t) + b_{\d^4 \delta}(t) \frac{k^4}{k_M^4} P_{11}(k;t,t)\,.
\label{eq:Phm_fin}
\end{align}
Note that, as expected, when $\lbrace b_{\delta,1} \rightarrow 1, ~  b_{\delta,2} \rightarrow 1, ~ b_{\delta,3} \rightarrow 1, ~ \
b_{\delta^2} \rightarrow 0, ~ b_{c_s}\rightarrow 0, ~ b_{\d^4 \delta} \rightarrow 0 \rbrace $, our formula reduces to the power spectrum of the standard dark matter case,
\be \label{eq:Pmm}
P_{\rm EFT-1-loop} = P_{11} + P_{1-\rm loop} - 2(2 \pi) c^2_{s(1)} \frac{k^2}{k_{\rm NL}^2} P_{11}\ .
\ee

Similarly, for the halo-halo power spectrum we have:
\begin{align}
 P_{hh}(k,t) =& b^2_{\delta,1}(t)\left( P_{11}(k;t,t) 
  +2\int \frac{d^3q}{(2\pi)^3} \;\left[\widehat{c}^{(2)}_{{\delta,1,s}}(\vec{k}-\q,\q)\right]^2P_{11}(q;t,t)P_{11}( |\vec{k}-\vec{q}|;t,t) \right. \non\\
  &\qquad\qquad\qquad\qquad\qquad\qquad \left.+6 \; P_{11}(k;t,t) \int \frac{d^3q}{(2\pi)^3} \;\left(\widehat{c}^{(3)}_{{\delta,1,s}}(\vec{k},-\q,\q) 
  + \sfrac{13}{63}\right)  P_{11}(q;t,t) \right) \non \\
  &+b_{\delta,1}(t) b_{\delta, 3}(t)\; 6\; P_{11}(\vec{k};t,t) \int \frac{d^3q}{(2\pi)^3} \;\left( \widehat{c}^{(3)}_{{\delta,3,s}}(\vec{k},-\q,\q) - \sfrac{47}{63}\right)  P_{11}(q;t,t) \non \\
  &+b^2_{\delta,2}(t) \; 2 \int \frac{d^3q}{(2\pi)^3} \;\left[F^{(2)}_s(\vec{k}-\vec{q},\vec{q}) - \widehat{c}^{(2)}_{{\delta,1,s}}(\vec{k}-\q,\q)\right]^2
                                                       P_{11}(q;t,t)P_{11}( |\vec{k}-\vec{q}|;t,t) \non \\
  &+b^2_{\delta^2}(t) \; 2 \int \frac{d^3q}{(2\pi)^3} \;P_{11}(q;t,t)P_{11}( |\vec{k}-\vec{q}|;t,t)\non \\
  &+b_{\delta,1}(t)b_{\delta, 2}(t)\; 4 \int \frac{d^3q}{(2\pi)^3} \;\widehat{c}^{(2)}_{{\delta,1,s}}(\vec{k}-\q,\q) \left(F^{(2)}_s(\vec{k}-\vec{q},\vec{q})\right.  \non \\
    &\qquad\qquad\qquad\qquad\qquad\qquad\qquad\qquad \left. - \widehat{c}^{(2)}_{{\delta,1,s}}(\vec{k}-\q,\q)\right) P_{11}(q;t,t)P_{11}( |\vec{k}-\vec{q}|;t,t)\non \\
  &+b_{\delta,1}(t)b_{\delta^2}(t)\; 4 \int \frac{d^3q}{(2\pi)^3} \;\widehat{c}^{(2)}_{{\delta,1,s}}(\vec{k}-\q,\q) P_{11}(q;t,t)P_{11}( |\vec{k}-\vec{q}|;t,t)\non \\
  &+b_{\delta,2}(t)b_{\delta^2}(t)\; 4 \int \frac{d^3q}{(2\pi)^3} \;
                                   \left(F^{(2)}_s(\vec{k}-\vec{q},\vec{q}) - \widehat{c}^{(2)}_{{\delta,1,s}}(\vec{k}-\q,\q)\right) P_{11}(q;t,t)P_{11}( |\vec{k}-\vec{q}|;t,t)\non \\
  &+b_{\delta,1}(t)b_{c_s}(t)\; 2 (-2(2\pi)) c^2_{s(1)}\frac{k^2}{k^2_{\text{NL}}}P_{11}(k;t,t) + 2 b_{\delta, 1}(t) b_{\d^4 \delta}(t) \frac{k^4}{k_M^4} P_{11}(k;t,t) \non \\
 &+ \text{Const}_{\epsilon} - 2 b_{\d^2 \epsilon}(t) \text{Const}_{\epsilon} \frac{k^2}{k_M^2} \non \\
& - 2\Sigma(t)^2 \left( b_{\delta,1}^2 + b_{\delta,2}^2 + b^2_{\delta^2} - 2 \; b_{\delta,1}b_{\delta,2}  
- 2 \; b_{\delta,1}b_{\delta^2} + 2 \; b_{\delta,2}b_{\delta^2} \right)   \, .
\label{eq:Phh_fin}
\end{align}
Again, in the limit $\lbrace b_{\delta,1} \rightarrow 1, ~  b_{\delta,2} \rightarrow 1, ~ b_{\delta,3} \rightarrow 1, ~ \
b_{\delta^2} \rightarrow 0, ~ b_{c_s}\rightarrow 0, ~ b_{\d^4 \delta} \rightarrow 0, ~ b_{\d^2 \epsilon} \rightarrow 0  \rbrace $, we find the dark matter power spectrum of eq.~\eqref{eq:Pmm}.

The renormalization procedure does not affect the three tree-level bispectra~\cite{Angulo:2015eqa}. Therefore, for each of the bias coefficients, we just have to perform the replacement (for $\tilde{c}_{\delta,1}$ and correspondingly for the rest):
 $ \tilde{c}_{\delta,1} \rightarrow \tilde{c}_{\delta,1,\text{ finite}} = b_{\delta_1} $. In this way, the same bias terms that appear in different observable can be given the same numerical value.

\subsection{IR-resummation}
\label{subsec:irresum}

It is only after renormalization that higher order terms in perturbation theory contribute with growing powers of $\epsilon_{s>}, \epsilon_{\delta<}$ or $\epsilon_{s<}$. Without this procedure, each order in perturbation theory would contribute equally, harming the convergence of the perturbative series.

As mentioned in the introduction, the parameter $\epsilon_{s<}$ is order one for the wavenumbers of interest, and its effect cannot be recovered by low-order Taylor expansion. While $\epsilon_{s<}$ cancels for IR-safe quantities, the parameter $\epsilon_{s_<}^{\rm safe}$ in (\ref{eq:epssafe}) does not, and it also give an order unity contribution to the Baryon Acoustic Oscillation (BAO) peak. It is therefore very useful to resum the contribution in $\epsilon_{s<}$ (or $\epsilon_{s_<}^{\rm safe}$). The way to do this was developed in~\cite{Senatore:2014via} by resumming the contributions that scale with this parameter~\footnote{The reason why a resummation converges to the correct result in this case is because the dependence of the final result on the parameter is analytic, as it can be clearly seen from the Lagrangian point of view~\cite{Senatore:2014via}. The analyticity is also at the base of the fact that the Taylor expansion converges, albeit at potentially an high order. This allows for some useful simplifications, for example when performing the IR-resummation in redshift space~\cite{Lewandowski:2015ziq}.}. Luckily, general relativity forbids a bias in the velocity between biased tracers and dark matter (see for example~\cite{Senatore:2014eva}). This implies that the same IR-resummation that is done for dark matter should apply, unaltered, for halos~\cite{Senatore:2014eva}. This prediction was verified in~\cite{Angulo:2015eqa}, which showed how this IR-resummation technique allows to correctly reproduce the BAO peak for halos. In our paper, we perform the same IR-resummation as in~\cite{Angulo:2015eqa}.

\section{Results and comparison to $N$-body simulations
\label{sec:results}}

In this Section we compare our theoretical results of Sec.~\ref{sec:equations} to $N$-body simulations. Since we are interested in how clustering of halos depends on the halo mass, the dark matter halo sample is divided into four 
subsamples (Bin0, 1, 2 and 3) according to their mass. Mass bins are defined as in~\cite{Okumura:2012xh, Vlah:2013lia}. Our goal is to show that by adding higher derivative bias terms, the predictions of the EFTofLSS done at the same order in the dark matter perturbations are similarly accurate for all mass bins. For this reason, we keep Bin0 with its seven original bias parameters \cite{Angulo:2015eqa} (i.e. only one higher derivative bias term in the power spectrum), and we add the higher derivative terms in the other bins with the hope to fit up to approximately the same $k$-reach as Bin0. Since Bin1 contains halos that are so light that it already performs comparably well to Bin0 by using no higher derivative bias, for this bin we use the same seven bias parameters as for Bin0 (i.e. we include no higher derivative bias for Bin1). For Bin2 and 3, instead, we fit for sixteen independent bias parameters; eight enter in the halo-halo auto power 
spectrum, six enter in the halo-matter cross power spectrum, eight enter in the halo-matter-matter cross bispectrum, 
and twelve in the halo-halo-matter and halo-halo-halo bispectra. 
We include below the results and corresponding plots obtained using these values. For more information about the details of the simulations, we refer to the former paper \cite{Angulo:2015eqa}, Sec.~3.1.

\subsection{Procedure for comparing to data}
\label{subsec:fitting}

For the two-point functions, we will not directly compare them to the data, but rather we will use the ratio between the halo-halo or halo-matter power spectrum and the matter-matter power spectrum, 
$$r^{(hh)}=P^{(hh)}/P^{(mm)}, \; r^{(hm)}=P^{(hm)}/P^{(mm)}.$$
This has the advantage of reducing the error in simulations due to sampling variance at low $k$'s~\cite{Seljak:2008xr, Seljak:2009af}. Instead, we will compare the bispectra directly, as the ratio for the bispectra does not cancel cosmic variance to an equal amount.

For convenience, we always use the data of the power-spectra up to $k= k_{\rm max,r} =0.277 h \, {\rm Mpc}^{-1}$. However for the bispectra, we choose a value $k_{\rm max,B}$ and we fit the results of Sec.~\ref{sec:equations} to every data points $B_{xyz}(k_1, k_2, k_3)$ such that ${\rm Max}[k_1, k_2, k_3] < k_{\rm max,B}$. We are interested in knowing the values of the bias coefficients depending on $k_{\rm max,B}$. As opposed to \cite{Angulo:2015eqa} where we redetermined the values of the bias coefficients for each $k_{\rm max,B}$, here we will fix their values using the following scheme:
\begin{itemize} \label{procedure}
\item For each $k_{\rm max,B}$, plot the value $b_i(k_{\rm max,B})$ of each bias coefficients with their 2-$\sigma$ error bars $\Delta b_i(k_{\rm max,B})$. The fit and the error bars are obtained using the Mathematica function {\it NonlinearModelFit}.

\item For each plot, determine the smallest value of $k$ such that it is impossible to draw a horizontal line from the vertical segment $[b_i(k) - \Delta b_i(k), \, b_i(k) + \Delta b_i(k)]$ to the far left of the plot without leaving the area delimited by the error bars. Call this value~$k_{\rm fit,i}$.

\item The final value of the coefficients $b_i$ is defined to be $b_i(k_{\rm fit})$, where $k_{\rm fit} = {\rm min}_j \, k_{\rm fit, j}$.
\end{itemize}

This procedure is practically the same as the one that has been implemented in~\cite{Foreman:2015lca,Cataneo:2016suz} for the dark matter power spectrum. It is designed to try to avoid the risk of overfitting and at the same time to adapt itself as much as possible to the errors in the simulations and in the theoretical predictions of the EFTofLSS. In fact, simulation data push the fit to higher wavenumber, because it is there that they have the smallest cosmic variance. However, the predictions of the EFTofLSS carry a theoretical error that grows at higher wavenumber. The procedure that we have explained just above determines the EFT coefficients at the highest possible value of $k_{\rm max}$ so that the numerical value that is obtained is compatible with the numerical values that are obtained using a lower~$k_{\rm max}$, where numerical data are affected by a larger uncertainty but the theoretical prediction by a smaller one. In this way, we enforce that in order to match higher wavenumbers, we are not degrading the match at low wavenumbers, where the theory should have a better match, but where simulations have bigger errors.

After the bias coefficients have been determined with the procedure that we have just described, it is possible to compute the $\chi^2$ and the $p$-value of the fit. The functions $\chi^2_{hm}$ and $\chi^2_{hh}$, which are polynomials of order six of six and eight bias coefficients respectively, are given by the formula:  
\begin{align}
 \chi^{2}_{r_{xy}} = \sum_{k_{i}\le k_{{\text{max}},r}}
 \frac{\left[r^{(xy)}(k_i)-r^{(xy)}_{\text{NL}}(k_{i})\right]^{2}}{\Delta r^{(xy)}_{\text{NL}}(k_{i})^{2}}\, , 
 \label{eq: chi2_00}
\end{align}
where $(xy)$ can mean either the halo-matter $(hm)$ cross-correlation or the halo-halo $(hh)$ auto-correlation. 
$k_i$ are all the points measured in the simulation, $r^{(xy)}$ is the ratio given by the theory and $r^{(xy)}_{\text{NL}}$ is the ratio given the numerical results of the simulation.
The error on the power spectrum ratios $\Delta r^{(xy)}_{\text{NL}}$ is detailed in \cite{Angulo:2015eqa}.
Similarly, one can construct the $\chi^2$ function for the three bispectra:  
\begin{align}
 \chi^{2}_{B_{hxy}} = \sum_{k_{1i},k_{2j},k_{3l}\le k_{{\text{max}},B}}
 \frac{\left[B^{hxy}(k_{1i},k_{2j},k_{3l}) -B^{hxy}_{\text{NL}}(k_{1i},k_{2j},k_{3l})\right]^{2}}{\Delta B^{hxy}_{\text{NL}}(k_{1i},k_{2j},k_{3l})^{2}}.
 \label{eq: chi2_Bk}
\end{align}
These three functions are polynomials of order six of eight bias coefficients for $B_{hmm}$ and of twelve bias coefficients for $B_{hhm}$ and $B_{hhh}$.
If we neglect all the correlations among the 
$\chi^2$ functions, the total $\chi^2_{\text{tot}}$ is a sum of all five components:
\begin{align}
\chi^2_{\text{tot}} = \chi^{2}_{r_{hm}} +\chi^{2}_{r_{hh}} +\chi^{2}_{B_{hmm}}+\chi^{2}_{B_{hhm}}+\chi^{2}_{B_{hhh}}.
\end{align}
The $p$-value is then computed using the formula 
\be \label{pvalue}
p = 1 - \frac{\gamma(K/2, K/2 * \chi^2_{\rm tot})}{\Gamma(K/2)},
\ee
where $K$ is the number of degrees of freedom, defined as the number of data points (which depends on $k_{\rm max, B}$) minus the number of bias coefficients, minus one. $\gamma( \cdot , \cdot)$ is the lower incomplete gamma function, and $\Gamma(\cdot)$ is the usual gamma function. Since both $K$ and $\chi^2_{\rm tot}$ depend on $k_{\rm max,B}$, $p$ does as well, and hence can be plotted as a function of $k_{\rm max,B}$.

\subsection{Results}
\label{subsec:fits}

\begin{figure*}[t!]
   \begin{center}
   \hspace*{-0.5cm}
   \includegraphics[scale=0.51]{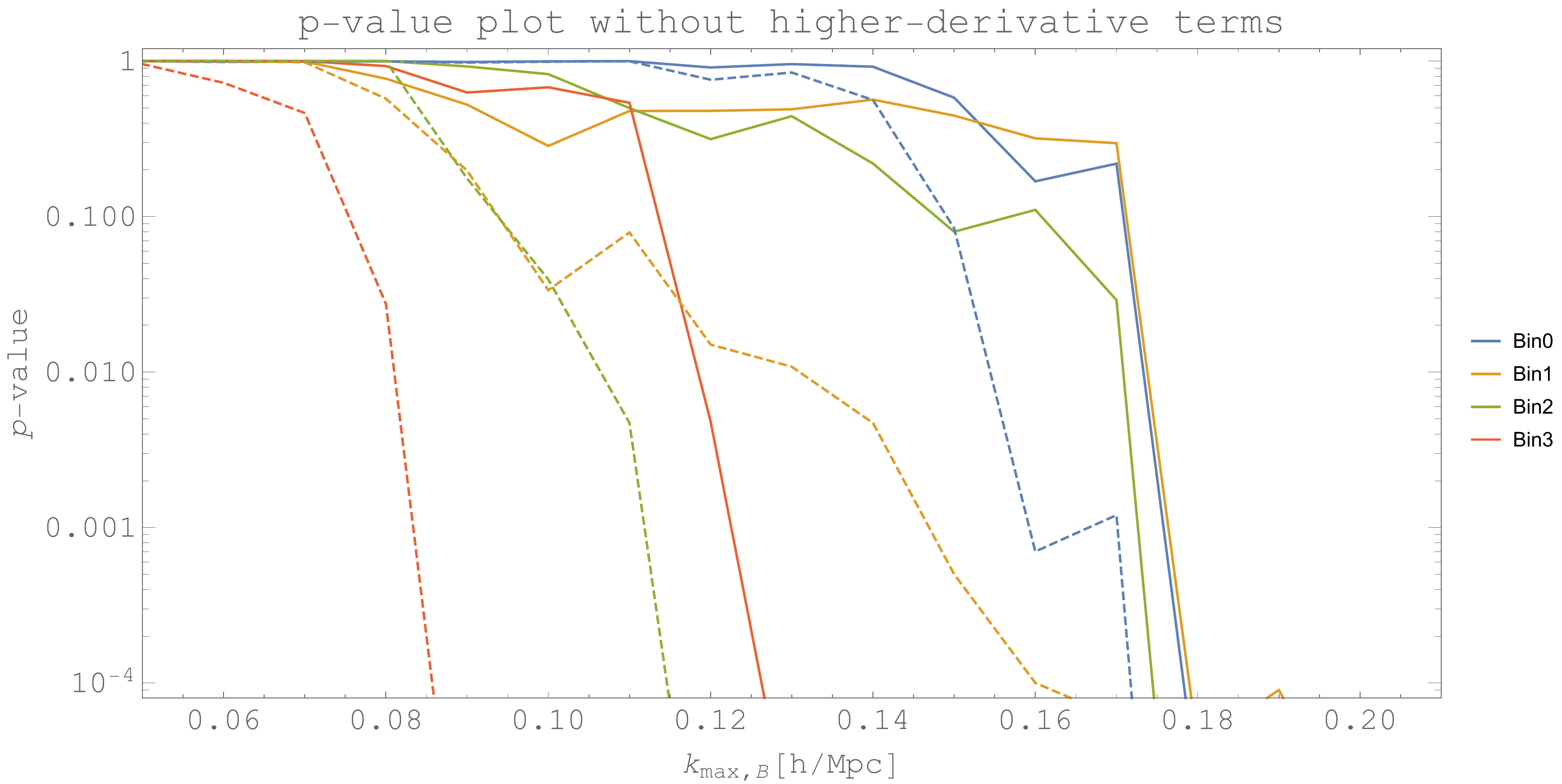}
   \end{center}
   \vspace*{-0.5cm}
   \caption{\small $p$-values of the fit as a function of $k_{\rm max,B}$ for each different bin, without any higher derivative term taken into account. The dashed lines represent the results that we would have gotten if we had not correct the typo described in Sec.~\ref{subsec:fits}, which was present in the numerical implementation of \cite{Angulo:2015eqa}. For the full lines, the $k$-reach is approximately the same for Bin0 and Bin1, and then becomes worse for Bin2 and Bin3, which motivates us to add higher derivative terms for those two bins. Our goal is to show that each bin acquires approximately the same $k$-reach as Bin0 and Bin1 just by adding higher derivative bias terms.}
   \label{fig:fitsevenparameters}
\end{figure*}

First, Fig.~\ref{fig:fitsevenparameters} is a plot of the $p$-value of the fits when only the seven bias parameters of \cite{Angulo:2015eqa} are taken into account (i.e. only the leading derivative bias in included in the two-point and three-point functions). There are two reasons why these plots are different from the corresponding ones in \cite{Angulo:2015eqa}. First, the values of the bias coefficients are not chosen the same way, see Sec.~\ref{subsec:fitting}, and second, a mistake in the Mathematica numerical implementation of the formulas of~\cite{Angulo:2015eqa} was corrected. This mistake was a factor 2 missing in the numerical implementation of eq.~\eqref{eq:Bhhh}~\footnote{In transcribing from the draft to the Mathematica notebook, it had been written as $\tilde{c}_{\delta,1}(t) \tilde{c}_{\epsilon \delta}(t)$ instead of $2 \tilde{c}_{\delta,1}(t) \tilde{c}_{\epsilon \delta}(t)$.}. 

If we define the $k_{\rm max}$ of the fit to be when the $p$-value becomes less than 0.1, we observe that Bin0 can be fitted up to $k_{\rm max} \simeq 0.17 h \, \rm Mpc^{-1}$. Furthermore Bin1 is just as good as Bin0, except for small values of $k$ where it is a bit worse. We therefore decide that we do not need to add any new terms to Bin1. Instead, Bin2 and especially Bin3 cannot be predicted up to the same $k_{\rm max}$ as Bin0, we therefore decide to add the higher derivative terms to these two bins, and just for these two. 

Notwithstanding the more conservative fitting procedure, overall we see that the $k$-reach of the fits are better than in \cite{Angulo:2015eqa}, which is due to the correction of the mistake stated above. This is a very encouraging sign, since the fact that the correction of a typo leads to an improvement of the result is the feature of a correct theory. It also tells us something about the remaining predicting power of the EFTofLSS. In the EFTofLSS, predictions depends on unknown bias parameters that need to be matched to the data. However, there are also terms that cannot be fudged. The fact that a correction of a mistake in the un-fudgeable terms leads to an improvement of the fit to the observations tells us that the un-fudgable terms carried with them a functional dependence that is different from the one of the bias coefficients and therefore it being wrong could not be compensated by adjusting the bias coefficients.  We conclude that the EFTofLSS maintains a lot of predicting power notwithstanding its free parameters.

\begin{figure*}[t!]
   \begin{center}
   \hspace*{-0.5cm}
   \includegraphics[scale=0.51]{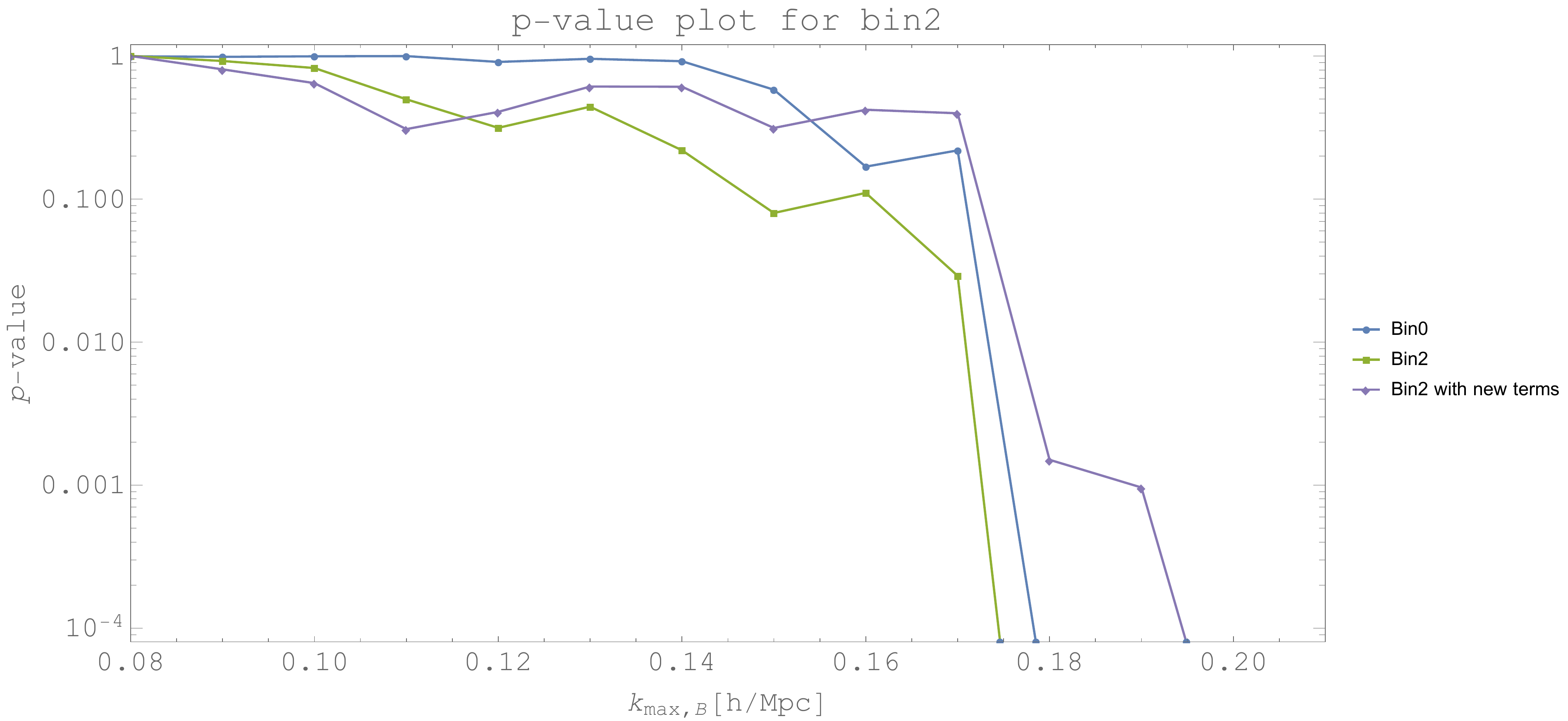}
   \end{center}
   \vspace*{-0.5cm}
   \caption{\small $p$-values of the fit for Bin2 as a function of $k_{\rm max,B}$, using the numerical values for the bias coefficients obtained at $k_{\rm fit}$. The blue line represents the curve of the $p$-values for Bin0, which defines the $k_{\rm max}$ that we want to achieve. We conclude that the addition of the higher derivative terms allows to fit Bin2 up to the same $k_{\rm max}$ as Bin0. }
   \label{fig:fitBin2}
\end{figure*}

After adding the new higher derivative bias parameters, the $k$-reach of Bin2 becomes slightly bigger than the one of Bin0, which achieves what we wished to show, see Fig.~\ref{fig:fitBin2}. Notice that the $p$-values at lower $k$ are not as good as without the higher derivative terms. This is probably due to a difficulty in our numerical procedure to determine the best fitting values when we have many bias coefficients (sixteen) to determine. Since this drop in the $p$-value at low $k$'s is quite marginal, we do not investigate it further.  
The result for Bin2 is however sufficiently good for our purposes.

\begin{figure*}[t!]
   \begin{center}
   \hspace*{-0.5cm}
   \includegraphics[scale=0.45]{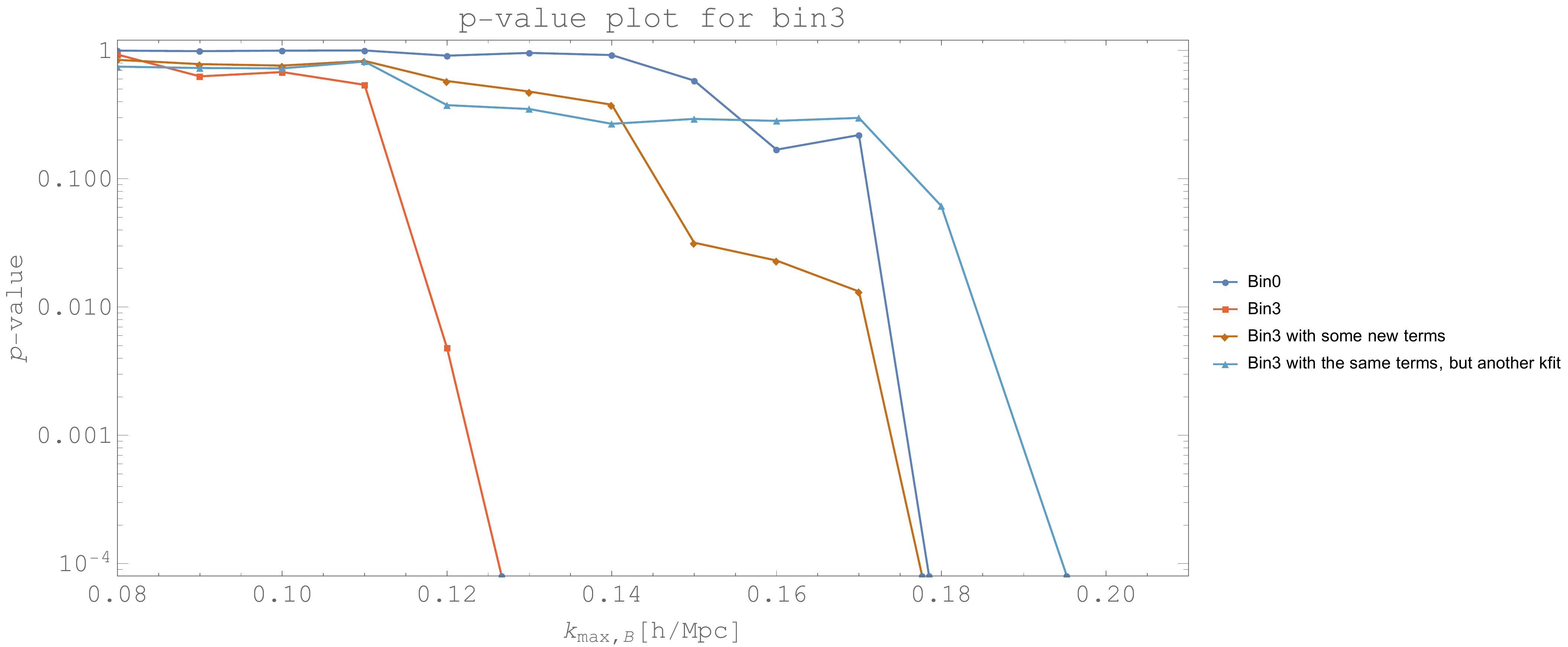}
   \end{center}
   \vspace*{-0.5cm}
   \caption{\small $p$-values of the fit for Bin3 as a function of $k_{\rm max,B}$, using the numerical values for the bias coefficients obtained at $k_{\rm fit}$. The blue line represents the curve of the $p$-values for Bin0, which defines the $k_{\rm max}$ that we want to approximately achieve. Since our numerical technique does not seem able to handle the large number of parameters that enter in the fit, we restrict ourselves to a subset of the higher derivative biases. The result is shown by the brown curve, which is quite close to the reach of Bin0. To give a sense if the slight under-reach of Bin3 is due to the residual theoretical error of the prediction or to our non-optimal numerical fitting procedure, in light blue we present the curve obtained by somewhat arbitrarily raising the $k_{\rm fit}$  to $0.15\hinvMpc$ from $0.12\hinvMpc$, which leads to a reach in $k$ as large as for Bin0. }
   \label{fig:fitBin3}
\end{figure*}

Finally, the results fro Bin3 are shown in Fig.~\ref{fig:fitBin3}. We find that our numerical implementation to minimizing the $\chi^2$ is not able to handle in this case the sixteen bias coefficients that needs to be fitted: as we increase the bias coefficients, the fit gets worse. For this reason, we restrict ourselves to the largest subset of higher derivative bias coefficients that gives a better fit than by using less parameters. When this is done, we find that the $k$-reach of Bin3 increases from $k \simeq 0.11 h \, \rm Mpc^{-1}$ to $k \simeq 0.14 h \, \rm Mpc^{-1}$. This is almost as good as the one of the other bins, which was our purpose. The mild underperformance of Bin3 with respect to Bin0, Bin1, and Bin2 could be due to the higher residual theoretical error that is present due to the fact that the bias coefficients for Bin3 are larger than for the other bins, as we argued in the Introduction. However, it could also be due to a difficulty in our numerical code in handling a large number of free parameters. In fact, even though the procedure described in Sec.~\ref{procedure} applied to Bin3 gives $k_{\rm fit} \simeq 0.12 h \, \rm Mpc^{-1}$, choosing arbitrarily $k_{\rm fit} \simeq 0.15 h \, \rm Mpc^{-1}$ gives a much better fit, with a larger $k$-reach than Bin0, and just mildly underperforming at some relatively low $k$'s. Given the goodness of the result, and the uncertainty in estimating the residual theoretical error, we conclude that is not worth to investigate the issue further, as it appears that the results for Bin3 sufficiently satisfy the theoretical expectations from the EFTofLSS. Similar considerations apply to $\langle [\d^2\epsilon]_{\bold k_1} [\epsilon]_{\bold k_2} [\epsilon]_{\bold k_3} \rangle'$, which we do not include for both Bin2 and Bin3, because it does not appear to improve the fit to the numerical data~\footnote{In particular, as it is the case when we include more higher derivative biases in Bin3, when we include this term (with a free coefficient), we find that the fit to data does not improve or becomes actually moderately worse. It is hard to explain why the fit can get moderately worse when adding a new term. One could argue that the $\d^2\epsilon$ contribution is relevant only at higher wavenumbers, so that we are allowed to include this term in the power spectrum and not in the bispectrum. Indeed, estimating its contribution from the numerical values we find in Table~\ref{tb:bias}, which are obtained assuming that this term is negligible in the bispectrum, one indeed finds that this term seems to be subleading in the bispectrum, while being relevant in the power spectrum. This justifies our procedure. But, even by adding a negligible term, one would expect the fit just not to improve, instead of getting worse, albeit moderately. Probably, it could be that our numerical fitting procedure has difficulties when we add this particular contribution, as it was the case for some terms in Bin3. Given that the main point of this paper is to show that by adding higher derivative terms the theory performs similarly well for all bins, which we achieve without adding this contribution, we decide to postpone further investigations on this issue to future work.}.

The plots of the values of the bias coefficients as a function of $k$, used for the determination of $k_{\rm fit}$ for each bin, as well as the plots of the ratio between the theoretical results and the numerical simulations results for the two-point functions, are given in App.~\ref{app:biasplot}. Table \ref{tb:bias} shows the values of the bias coefficients found by our numerical fitting procedure
when applied to the combination of the five observables: two power spectra and three bispectra. We can see that the error bars for almost each bias coefficient are relatively large. But we have to take into account that the coefficients are correlated between one another, so we provide the correlation matrix for each bin in App.~\ref{app:biascorrelation}. We see that there is a relatively large correlation at least among some of the bias coefficients (see~\cite{Foreman:2015lca} for a discussion how these correlations can be anticipated in the context of the EFTofLSS.).

\begin{table*}[t!]
\caption{Best fit bias parameters table for the four bins, measured in units of $\hinvMpc$ to the relevant power. Bin0 and Bin1 do not need any higher derivative term to be fitted up to the $k_{\rm max}$ of interest. Bin2 is fitted using the seven original bias parameters plus the nine leading higher derivative terms. Bin3 would also in principle need the nine leading higher derivative terms but the numerical fitting procedure we use could not efficiently handle so many parameters so we forced four of them to be zero. Error bars on some bias coefficients are large, but they are quite correlated (see the correlation matrices in App.~\ref{app:biascorrelation}).}
\centering 
\setlength{\tabcolsep}{8pt}
\renewcommand{\arraystretch}{1.0}
\begin{tabular}{c|cccccc}
\hline\hline
  & Bin0 & Bin1 & Bin2 & Bin3 \\ [0.5ex] 
\hline
$b_{\df,1}$  & $  1.00\pm0.01$ & {$1.30\pm0.01$} & $1.88\pm0.03$ & $2.91\pm0.5$ \\
$b_{\df,2}$  & $  0.34\pm0.26$ &  {$0.10\pm0.33$} & $-1.68\pm1.88$ & $-3.12\pm1.85$  \\
$b_{\df,3}$  & $  0.37\pm0.48$ &  {$0.20\pm0.68$} & $1.29\pm2.80$ & $2.83\pm2.48$  \\
$b_{\df^2}$ & $  0.17\pm0.22$ &  {$0.64\pm0.30$} & $2.74\pm1.60$ & $5.35\pm1.73$  \\
$b_{c_s}$     & $  0.64\pm0.60$ &  {$0.67\pm0.83$} & $-1.12\pm6.03$ & $-6.36\pm3.15$  \\
$b_{\epsilon \df}$ &  $  0.24\pm0.07$ &  {$0.47\pm0.07$} & $1.30\pm0.15$ & $1.88\pm0.03$  \\
${\rm Const}_\epsilon$   &  $5600\pm400$ &  {$11604\pm567$} & $31535\pm3506$ & $123385\pm4543$  \\
$b_{\d^2 \delta,1}/k_m^2$ & - & - & $-28.4\pm23$ & - \\
$b_{\d^2 \delta,2}/k_m^2$ & - & - & $-31.2\pm30.4$ & - \\
$b_{\d^2 \delta^2}/k_m^2$ & - & - & $41.3\pm39.2$ & $99.39\pm37.6$ \\
$b_{\d^2 s^2}/k_m^2$ & - & - & $59.2\pm47.4$ & - \\
$b_{(\d \delta)^2/k_m^2}$ & - & - & $-39.4\pm24.7$ & $-168.55\pm45.79$ \\
$b_{\d^4 \delta}/k_m^4$ & - & - & $-12.20\pm87.65$ & - \\
$b_{\d^2 \epsilon}/k_m^2$  & - & - & $0.83\pm0.93$ & $0.26\pm 0.66$ \\
$b_{\d^2 \epsilon \delta}/k_m^2$ & - & - & $5.15\pm15.9$ & $-4.07\pm12.67$ \\
$b_{\epsilon \d^2 \delta}/k_m^2$ ñ& - & - & $-3.9\pm25.1$ & $53.26\pm24.29$ \\
\hline
\end{tabular}
\label{tb:bias}
\end{table*}


\section{Conclusions}
\label{sec:conclusion}

We have argued that The Effective Field Theory of Cosmological Large Scale Structures (EFTofLSS) when applied to biased tracers predicts the following. If we compute a correlation function for biased tracers at a given order in the dark matter non-linearities, the theoretical error is larger for tracers with larger biases. This theoretical error is mainly controlled by the size of the higher derivative terms, and by a subleading correction due to the size of the bias coefficients. Therefore, a prediction of the EFTofLSS is that if all tracers are treated as equal, the predictions for highly biased tracers underperform with respect to the ones for less biased ones. However, by adding just higher derivative operators, it is predicted  that the theoretical results for all traces should work comparably well. We have implemented this construction by adding the contribution from higher derivative biases just for highly biased tracers, and have found that indeed all tracers perform comparably well. At the order at which we have computed, we are able to predict all two-point and three-point functions up to $k\sim 0.17\hinvMpc$ which is a remarkable improvement with respect to former techniques.

Of course, our findings are affected by the precision of our numerical data, and by the relatively low order of the calculations, that do not allow us to use data at very high wavenumbers. It will therefore be very interesting to perform a similar comparison by performing higher-order calculations, by computing higher-$N$ point functions, as well as by using more precise numerical data and potentially more accurate fitting procedures, so that our findings can be better verified. We plan to do this in future work.

\section*{Acknowledgments}

We thank Mikhail (Misha) Shaposhnikov for support. V.M. thanks the Stanford Institute for Theoretical Physics for hospitality.
L.S. is partially supported by DOE Early Career Award DE-FG02-12ER41854. Z.V. is supported in part by the U.S. Department of Energy contract to SLAC no. DEAC02-76SF00515.



\appendix
\section*{Appendix}

\section{$\mathbb{C}_i$ operators}
\label{app:operators}

In this Appendix we provide some useful formulas to complete the expressions that we provide in the main part of the paper. In particular, we give the definition of the operators of eq.~\eqref{eq:CoI_all}, the definition of the kernels of eq.~\eqref{eq:CoI} and the different degeneracy relations. 

First, here are the definition of the operators $\mathbb{C}_i^{(n)}$ of eq.~\eqref{eq:CoI_all} in terms of the quantities in eq.~\eqref{eq:euler_bias_4}:
\bea \label{eq:C_all_explicit}
&&\text{1st order:}  \\ \nonumber
&&\mathbb{C}^{(1)}_{\delta,1} (\vec k,t) = \delta^{(1)} (\vec k,t)\, ,   \\ \nonumber
&& {\mathbb{C}_{\epsilon}(\vec k,t)= [\epsilon]_{\vec k}  }  \, , \\ \nonumber
&& {\mathbb{C}^{(1)}_{\epsilon\delta}(\vec k,t)= [\epsilon\delta]^{(1)}_{\vec k} } \, ,\\ \nonumber
&& \\ \nonumber
&&\text{2nd order:}\\ \nonumber
&&\mathbb{C}^{(2)}_{\delta,1} (\vec k,t) = [\d_i \delta^{(1)}\; \frac{\d^i}{\d^2}\theta^{(1)}]_{\vec k}(t)\,  , \\ \nonumber
&&\mathbb{C}^{(2)}_{\delta,2} (\vec k,t) = \delta^{(2)}(\vec k,t) - [\d_i \delta^{(1)}\; \frac{\d^i}{\d^2}\theta^{(1)}]_{\vec k}(t) \, ,\\ \nonumber
&&\mathbb{C}^{(2)}_{\delta^2,1} (\vec k,t) =    [\delta^2]_{\vec k}^{(2)}(t) \, ,\\ \nonumber
&&\mathbb{C}^{(2)}_{s^2,1}(\vec k,t) =   [s^2]_{\vec k}^{(2)}(t)  \,  ,\\ \nonumber
&&\mathbb{C}^{(2)}_{\epsilon,2} (\vec k,t) = [\epsilon^{(2)}]_{\vec k} \, , \\ \nonumber
&& \\ \nonumber
&&\text{3rd order:}\\ \nonumber
&&\mathbb{C}^{(3)}_{\delta,1} (\vec k,t) = \frac{1}{2}  [\d_i \delta^{(1)}\; \frac{\d^i}{\d^2}\theta^{(2)}]_{\vec k}(t) 
+\frac{1}{2}  \left( [\d_i \delta^{(1)}\; \frac{\d_j\d^i}{\d^2}\theta^{(1)}\; \frac{\d^j}{\d^2}\theta^{(1)}]_{\vec k}(t) \right.  \\ \nonumber
&&\qquad\qquad\qquad\qquad\qquad\qquad\qquad\qquad\qquad\qquad \left.
+[\d_i\d_j \delta^{(1)}\;\frac{\d^i}{\d^2}\theta^{(1)}\frac{\d^j}{\d^2}\theta^{(1)}]_{\vec k}(t)\right)  \, ,\\  \nonumber
&&\mathbb{C}^{(3)}_{\delta,2} (\vec k,t) =  [\d_i \delta^{(2)}\; \frac{\d^i}{\d^2}\theta^{(1)}]_{\vec k}(t)  
- \left( [\d_i \delta^{(1)}\; \frac{\d_j\d^i}{\d^2}\theta^{(1)}\; \frac{\d^j}{\d^2}\theta^{(1)}]_{\vec k}(t) \right. \\  \nonumber
&&\qquad\qquad\qquad\qquad\qquad\qquad\qquad\qquad\qquad\qquad \left.
+[\d_i\d_j \delta^{(1)}\;\frac{\d^i}{\d^2}\theta^{(1)}\frac{\d^j}{\d^2}\theta^{(1)}]_{\vec k}(t)\right)   \, ,\\ \nonumber
&&\mathbb{C}^{(3)}_{\delta,3} (\vec k,t) =  \delta^{(3)}(\vec k,t) - [\d_i \delta^{(2)}\; \frac{\d^i}{\d^2}\theta^{(1)}]_{\vec k}(t) 
+ \frac{1}{2} \left( [\d_i \delta^{(1)}\; \frac{\d_j\d^i}{\d^2}\theta^{(1)}\; \frac{\d^j}{\d^2}\theta^{(1)}]_{\vec k}(t) \right. \\  \nonumber
&&\qquad\qquad\qquad\qquad\qquad\qquad\qquad\qquad\qquad\qquad \left.
+[\d_i\d_j \delta^{(1)}\;\frac{\d^i}{\d^2}\theta^{(1)}\frac{\d^j}{\d^2}\theta^{(1)}]_{\vec k}(t)\right)   \, , \\ \nonumber
&&\mathbb{C}^{(3)}_{\delta^2,1} (\vec k,t) = 2 [\delta^{(1)} \d_i \delta^{(1)} \frac{\d^i}{\d^2} \theta^{(1)}]_{\vec k}(t)  \, ,\\ \nonumber 
&&\mathbb{C}^{(3)}_{\delta^2,2} (\vec k,t) = [\delta^2]^{(3)}_{\vec k}(t)-2  [\delta^{(1)} \d_i \delta^{(1)} \frac{\d^i}{\d^2} \theta^{(1)}]_{\vec k}(t)     \, ,\\ \nonumber 
&&\mathbb{C}^{(3)}_{\delta^3,1} (\vec k,t) = [\delta^3]^{(3)}_{\vec k}(t) \,  ,\\ \nonumber 
&&\mathbb{C}^{(3)}_{s^2,1} (\vec k,t) = 2 [s_{lm}^{(1)}\d_i (s^{lm})^{(1)}\frac{\d^i}{\d^2}\theta^{(1)}]_{\vec k}(t) \, , \\ \nonumber 
&&\mathbb{C}^{(3)}_{s^2,2} (\vec k,t) = [s^2]^{(3)}_{\vec k}(t)-2  [s_{lm}^{(1)}\d_i (s^{lm})^{(1)}\frac{\d^i}{\d^2}\theta^{(1)}]_{\vec k}(t)    \,  ,\\ \nonumber 
&&\mathbb{C}^{(3)}_{s^3,1} (\vec k,t) = [s^3]^{(3)}_{\vec k}(t)   \,  , \\ \nonumber 
&&\mathbb{C}^{(3)}_{st,1} (\vec k,t) = [st]^{(3)}_{\vec k}(t) \, ,\\ \nonumber 
&&\mathbb{C}^{(3)}_{\psi,1} (\vec k,t) = \psi^{(3)}(\vec k,t)   \,  ,\\ \nonumber 
&&\mathbb{C}^{(3)}_{\delta s^2,1} (\vec k,t) =   [\delta s^2]^{(3)}_{\vec k}(t) \, , \\  \nonumber
&&{ \mathbb{C}^{(3)}_{\delta,3_{c_s}} (\vec k,t) = \delta^{(3)}_{c_s}(\vec k,t)\, , } \\ \nonumber
&& \\ \nonumber
&&\text{Higher derivatives:}\\ \nonumber
&&\mathbb{C}^{(1)}_{\d^2 \delta, 1}(\bold k, t) = - \frac{k^2}{k_M^2} \delta^{(1)}(\bold k, t) \, , \\ \nonumber
&&\mathbb{C}^{(1)}_{\d^4 \delta}(\bold k, t) = \frac{k^4}{k_M^4} \delta^{(1)}(\bold k, t) \, , \\ \nonumber
&&\mathbb{C}^{(2)}_{\d^2 \delta, 1}(\bold k, t) = [\frac{\d^2}{k_M^2} \d_i \delta^{(1)} \frac{\d^i}{\d^2} \theta^{(1)}]_{\bold k}(t) \, , \\ \nonumber
&&\mathbb{C}^{(2)}_{\d^2 \delta, 2}(\bold k, t) = - \frac{k^2}{k_M^2} \delta^{(2)}(\bold k, t) - [\frac{\d^2}{k_M^2} \d_i \delta^{(1)} \frac{\d^i}{\d^2} \theta^{(1)}]_{\bold k}(t) \, , \\ \nonumber
&&\mathbb{C}^{(2)}_{\d^2 \delta^2}(\bold k, t) =  - \frac{k^2}{k_M^2} [\delta^2]^{(2)}(\bold k, t) \, , \\ \nonumber
&&\mathbb{C}^{(2)}_{\d^2 s^2}(\bold k, t) =  - \frac{k^2}{k_M^2} [s^2]^{(2)}(\bold k, t) \, , \\ \nonumber
&&\mathbb{C}^{(2)}_{(\d \delta)^2}(\bold k, t) =  [\frac{\d_i}{k_M} \delta^{(1)} \frac{\d^i}{k_M} \delta^{(1)}]_{\bold k}(t) \, , \\ \nonumber
&&\mathbb{C}_{\d^2 \epsilon}(\bold k, t) =  - \frac{k^2}{k_M^2} [\epsilon]_{\bold k}(t) \, , \\ \nonumber
&&\mathbb{C}^{(1)}_{\d^2 \epsilon \delta}(\bold k, t) = [(\frac{\d^2}{k_M^2} \epsilon) \delta]^{(1)}_{\bold k}(t) \, , \\ \nonumber
&&\mathbb{C}^{(1)}_{\epsilon \d^2 \delta}(\bold k, t) = [\epsilon \frac{\d^2}{k_M^2} \delta]^{(1)}_{\bold k}(t) \, .
\eea 

The kernels of eq.~\eqref{eq:CoI} are defined from the operators $\mathbb{C}_i^{(n)}$ according to the following formula :
\begin{align} \label{eq:C_all_explicit_2}
\mathbb{C}^{(1)}_i (\vec k,t)&= \int \frac{d^3q_1}{(2\pi)^3\;)}\widehat{c}^{(1)}_{s,i} (q_1)\delta_D^{(3)}(\vec k-\vec q_1)\delta^{(1)} (\vec q_1,t), \\ \nonumber
\mathbb{C}^{(2)}_i (\vec k,t)&= \int \frac{d^3q_1}{(2\pi)^3}\;\frac{d^3q_2}{(2\pi)^3}\;\; 
                                           \widehat{c}^{(2)}_{s,i} (q_1,q_2)\delta_D^{(3)}(\vec k-\vec q_1-\vec q_2)\delta^{(1)} (\vec q_1,t)\delta^{(1)} (\vec q_2,t), \\ \nonumber
\mathbb{C}^{(3)}_i (\vec k,t)&= \int \frac{d^3q_1}{(2\pi)^3}\;\frac{d^3q_2}{(2\pi)^3}\;\frac{d^3q_3}{(2\pi)^3}\;\; 
                                           \widehat{c}^{(3)}_{s,i} (\vec q_1,\vec q_2,\vec q_3)
                                           \delta_D^{(3)}(\vec k-\vec q_1-\vec q_2-\vec q_3)\delta^{(1)} (\vec q_1,t)\delta^{(1)} (\vec q_2,t)\delta^{(1)} (\vec q_3,t), \\ \nonumber
\end{align}
The lower index $_s$ means that the kernel is symmetrized, that is to say averaged over the sum of all possible permutations of the variables. Formula \eqref{eq:C_all_explicit_2} yields the following kernels, in their unsymmetrized version:
\bea \label{eq:c_all_explicit}
\text{1st order:} \\ \nonumber
\widehat{c}^{(1)}_{\delta,1}(\vec q_1) &= & 1, \\ \nonumber
\\ \nonumber
\text{2nd order:}\\ \nonumber
\widehat{c}^{(2)}_{\delta,1}(\vec q_1,\vec q_2) &=& \sfrac{\V q_1\cdot\V q_2 }{q_1^2}, \\ \nonumber
\widehat{c}^{(2)}_{\delta,2}(\vec q_1,\vec q_2) &=& F^{(2)}\left(\V q_1, \V q_2\right) -  \sfrac{\V q_1\cdot\V q_2 }{q_1^2},\\ \nonumber
\widehat{c}^{(2)}_{\delta^2,1} (\vec q_1,\vec q_2) &=& 1,\\ \nonumber
\widehat{c}^{(2)}_{s^2,1} (\vec q_1,\vec q_2) &=& \sfrac{(\V q_1\cdot\V q_2)^2}{q_1^2 q_2^2} - \sfrac{1}{3},\\ \nonumber
\nonumber\\
\text{3rd order:}\nonumber \\ \nonumber 
\widehat{c}^{(3)}_{\delta,1}(\vec q_1,\vec q_2,\vec q_3) &=& \sfrac{1}{2} \left(\sfrac{\left(\V q_1 \cdot \V q_2+\V q_1 \cdot \V q_3\right)}{q_2^2+q_3^2
                                                                     +2 \V q_2 \cdot \V q_3} G^{(2)}\left(\V q_2, \V q_3\right)
                                                                     +\sfrac{\V q_1 \cdot \V q_2 \left(\V q_1 \cdot \V q_3
                                                                     + \V q_2\cdot \V q_3\right)}{q_2^2 q_3^2}\right), \\ \nonumber
\widehat{c}^{(3)}_{\delta,2}(\vec q_1,\vec q_2,\vec q_3) &=& \sfrac{\left(\V q_1\cdot \V q_3+\V q_2 \cdot \V q_3\right) }{q_2^2 q_3^2}
                                                                \left(F^{(2)}\left(\V q_1, \V q_2\right) q_2^2-\V q_1 \cdot \V q_2 \right), \\ \nonumber
\widehat{c}^{(3)}_{\delta,3}(\vec q_1,\vec q_2,\vec q_3) &=& F^{(3)}\left(\V q_1, \V q_2,  \V q_3\right)+\sfrac{\left(\V q_1+ \V q_2\right)\cdot \V q_3}{2q_2^2 q_3^2}
                                                                 \left(\V q_1\cdot \V q_2-2  F^{(2)}\left(\V q_1, \V q_2\right) q_2^2\right)\\ \non
                                                                 &&\qquad\qquad\qquad\qquad\qquad
                                                                 -\sfrac{\V q_1\cdot \left(\V q_2+ \V q_3\right)}{2(q_2^2+q_3^2 + 2 \V q_2\cdot \V q_3)} G^{(2)}\left(\V q_2, \V q_3\right) \\ \non
\widehat{c}^{(3)}_{\delta^2,1} (\vec q_1,\vec q_2,\vec q_3) &=& 2\sfrac{\V q_2 \cdot \V q_3}{q_3^2} \\ \nonumber 
\widehat{c}^{(3)}_{\delta^2,2} (\vec q_1,\vec q_2,\vec q_3) &=& 2 F^{(2)}\left(\V q_1, \V q_2\right) - 2\sfrac{\V q_2 \cdot \V q_3}{q_3^2}  \\ \nonumber 
\widehat{c}^{(3)}_{\delta^3,1} (\vec q_1,\vec q_2,\vec q_3) &=& 1 \\ \nonumber 
\widehat{c}^{(3)}_{s^2,1} (\vec q_1,\vec q_2,\vec q_3) &=& 2\sfrac{\V q_2 \cdot \V q_3}{q_3^2} 
                                                                  \left( \sfrac{\left(\V q_1 \cdot \V q_2\right)^2}{q_1^2 q_2^2} -\sfrac{1}{3} \right) \\ \nonumber 
\widehat{c}^{(3)}_{s^2,2} (\vec q_1,\vec q_2,\vec q_3) &=& 2F^{(2)}\left(\V q_1, \V q_2\right)\left( \sfrac{\left((\V q_1 + \V q_2)\cdot \V q_3\right)^2}
                                                                  {(\V q_1 + \V q_2)^2 q_3^2} -\sfrac{1}{3} \right)-2\sfrac{\V q_2 \cdot \V q_3}{q_3^2} 
                                                                  \left( \sfrac{\left(\V q_1 \cdot \V q_2\right)^2}{q_1^2 q_2^2} -\sfrac{1}{3} \right)\\ \nonumber 
\widehat{c}^{(3)}_{s^3,1} (\vec q_1,\vec q_2,\vec q_3) &=& \sfrac{9 \V q_1\cdot \V q_2 \V q_1\cdot \V q_3 \V q_2\cdot \V q_3-3 \left(\V q_1\cdot \V q_3\right)^2 q_2^2-3 
                                                                \left(\V q_1\cdot \V q_2\right)^2 q_3^2 
                                                                + q_1^2 \left( -3 \left(\V q_2\cdot \V q_3\right)^2+2 q_2^2 q_3^2\right)}{9 q_1^2 q_2^2 q_3^2} \\ \nonumber 
\widehat{c}^{(3)}_{st,1} (\vec q_1,\vec q_2,\vec q_3) &=& \left( G^{(2)}\left(\V q_1, \V q_2\right) - F^{(2)}\left(\V q_1, \V q_2\right) \right) 
                                                                 \left( \sfrac{\left( \left(\V q_1 + \V q_2 \right) \cdot \V q_3 \right)^2}{(\V q_1^2 + \V q_2)^2 q_3^2} -\sfrac{1}{3} \right)\\ \nonumber 
\widehat{c}^{(3)}_{\psi,1} (\vec q_1,\vec q_2,\vec q_3) &=&   G^{(3)}\left(\V q_1, \V q_2, \V q_3\right) - F^{(3)}\left(\V q_1, \V q_2, \V q_3\right)  \\ \nonumber 
                                                                  &&+ 2 F^{(2)}\left(\V q_1, \V q_2\right) \left( F^{(2)}\left(\V q_1 + \V q_2, \V q_3\right) 
                                                                  - G^{(2)}\left(\V q_1+ \V q_2, \V q_3\right)  \right) \\ \nonumber 
\widehat{c}^{(3)}_{\delta s^2,1} (\vec q_1,\vec q_2,\vec q_3) &=& \sfrac{\left(\V q_1 \cdot \V q_2\right)^2}{q_1^2 q_2^2} -\sfrac{1}{3}  \\ \nonumber 
\\ \nonumber
\text{Higher derivative terms:}\\ \nonumber
\widehat{c}^{(1)}_{\partial^2 \delta, 1}(\bold q_1) & = & -\frac{q_1^2}{k_M^2}, \nonumber
\\
\widehat{c}^{(1)}_{\partial^4 \delta}(\bold q_1) & = & \frac{q_1^4}{k_M^4}, \nonumber
\\
\widehat{c}^{(2)}_{\partial^2 \delta, 1}(\bold q_1, \bold q_2) & = & - \frac{q_1^2}{k_M^2} \frac{\bold q_1 \cdot \bold q_2}{q_2^2}, \nonumber
\\
\widehat{c}^{(2)}_{\partial^2 \delta, 2}(\bold q_1, \bold q_2) & = & -\frac{(\bold q_1 + \bold q_2)^2}{k_M^2} F^{(2)}(\bold q_1, \bold q_2) + \frac{q_1^2}{k_M^2} \frac{\bold q_1 \cdot \bold q_2}{q_2^2}, \nonumber
\\
\widehat{c}^{(2)}_{\partial^2 \delta^2}(\bold q_1, \bold q_2) & = & -\frac{(\bold q_1 + \bold q_2)^2}{k_M^2}, \nonumber
\\
\widehat{c}^{(2)}_{\partial^2 s^2}(\bold q_1, \bold q_2) & = & -\frac{(\bold q_1 + \bold q_2)^2}{k_M^2} \left( \frac{(\bold q_1 \cdot \bold q_2)^2}{q_1^2 q_2^2} - \frac{1}{3} \right), \nonumber
\\
\widehat{c}^{(2)}_{(\partial \delta)^2}(\bold q_1, \bold q_2) & = & -\frac{\bold q_1 \cdot \bold q_2}{k_M^2}. \nonumber
\eea

Now, here are the linear combinations that relate the degenerate operators in \eqref{eq:C_all_explicit} to the {\it BoD} basis defined in Sec.~\ref{subsec:field}:
\\
\bea \label{eq:C_all_explicit_4}
\text{2nd order operators:} \nonumber
\\
\mathbb{C}^{(2)}_{s^2,1} & = & \sfrac{7}{2} \mathbb{C}^{(2)}_{\delta, 2} -  \sfrac{17}{6} \mathbb{C}^{(2)}_{\delta^2, 1}, \nonumber
\\
&&\\ \nonumber
\text{3rd order operators:} \nonumber 
\\
\mathbb{C}^{(3)}_{s^2, 1} & = &  \sfrac{7}{2} \mathbb{C}^{(3)}_{\delta,2} - \sfrac{17}{6} \mathbb{C}^{(3)}_{\delta^2, 1}, \nonumber
\\
\mathbb{C}^{(3)}_{s^3, 1} & = & \sfrac{45}{4}\mathbb{C}^{(3)}_{\delta, 3} -\sfrac{137}{16} \mathbb{C}^{(3)}_{\delta^2, 2} + \sfrac{511}{72} \mathbb{C}^{(3)}_{\delta^3, 1} -\sfrac{3}{4} \mathbb{C}^{(3)}_{s^2, 2}, \nonumber
\\
\mathbb{C}^{(3)}_{st, 1} & = & \sfrac{9}{2} \mathbb{C}^{(3)}_{\delta, 3} -\sfrac{71}{24} \mathbb{C}^{(3)}_{\delta^2, 2} + \sfrac{25}{12} \mathbb{C}^{(3)}_{\delta^3, 1} - \sfrac{1}{2} \mathbb{C}^{(3)}_{s^2, 2}, \nonumber
\\
\mathbb{C}^{(3)}_{\psi, 1} & = & 2 \mathbb{C}^{(3)}_{\delta,3} - \sfrac{55}{42}\mathbb{C}^{(3)}_{\delta^2, 2} + \mathbb{C}^{(3)}_{\delta^3, 1} - \sfrac{2}{7}\mathbb{C}^{(3)}_{s^2, 2}, \nonumber
\\
\mathbb{C}^{(3)}_{\delta s^2, 1} & = & \sfrac{7}{4} \mathbb{C}^{(3)}_{\delta^2, 2} - \sfrac{17}{6} \mathbb{C}^{(3)}_{\delta^3, 1}, \nonumber
\\
&&\\ \nonumber
\text{Higher derivative terms :} \label{eq:higher_deri}\nonumber
\\
\mathbb{C}^{(1)}_{\d^2 \delta, 1} & = & \sfrac{1}{2 \pi} \sfrac{1}{c^2_{s(1)}} \sfrac{k_{NL}^2}{k_M^2} \mathbb{C}^{(3)}_{\delta, 3_{cs}}, \nonumber
\\
\mathbb{C}^{(1)}_{\d_i \d_j s^{ij}, 1} & = & \sfrac{2}{3} \mathbb{C}^{(1)}_{\d^2 \delta, 1}, \nonumber
\\
\mathbb{C}^{(2)}_{\d_i \d_j s^{ij}, 1} & = & - \sfrac{2}{3} \mathbb{C}^{(2)}_{\d^2 \delta, 1} + \mathbb{C}^{(2)}_{\d^2 \delta, 2} - \sfrac{17}{21} \mathbb{C}^{(2)}_{\d^2 \delta^2} - \sfrac{2}{7} \mathbb{C}^{(2)}_{\d^2 s^2} + \mathbb{C}^{(2)}_{(\d \delta)^2}, \nonumber
\\
\mathbb{C}^{(2)}_{\d_i \d_j s^{ij}, 2} & = & \sfrac{4}{3} \mathbb{C}^{(2)}_{\d^2 \delta, 1} - \sfrac{1}{3} \mathbb{C}^{(2)}_{\d^2 \delta, 2} + \sfrac{17}{21} \mathbb{C}^{(2)}_{\d^2 \delta^2} + \sfrac{2}{7} \mathbb{C}^{(2)}_{\d^2 s^2} - \mathbb{C}^{(2)}_{(\d \delta)^2}, \nonumber
\\
\mathbb{C}^{(2)}_{\d_i \d_j \d_k \phi \, \d^i \d^j \d^k \phi} = \mathbb{C}^{(2)}_{\d_i \d_j v_k \, \d^i \d^j v^k} = \mathbb{C}^{(2)}_{\d_i \d_j v_k \, \d^i \d^k v^j} & = & \sfrac{1}{2} \mathbb{C}^{(2)}_{\d^2 \delta, 2} - \sfrac{4}{7} \mathbb{C}^{(2)}_{\d^2 \delta^2} - \sfrac{9}{14} \mathbb{C}^{(2)}_{\d^2 s^2} + \sfrac{1}{2} \mathbb{C}^{(2)}_{(\d \delta)^2}, \nonumber
\\
\mathbb{C}^{(2)}_{\d_i \d_j t^{ij}} & = & - \sfrac{8}{63} \mathbb{C}^{(2)}_{\d^2 \delta^2} + \sfrac{4}{21} \mathbb{C}^{(2)}_{\d^2 s^2}.
\eea 

Finally, here are the definitions of the new bias coefficients $\tilde{c}_i$ that appear in equation \eqref{eq:CoI}, with respect to the old ones: 
\bea \label{eq:c_tildas}
\tilde{c}_{\delta, 1} & = & c_{\delta, 1}, \nonumber
\\
\tilde{c}_{\delta, 2} & = & c_{\delta, 2} + \sfrac{7}{2} c_{s^2, 1}, \nonumber
\\
\tilde{c}_{\delta, 3} & = & c_{\delta, 3} + \sfrac{45}{4} c_{s^3} + \sfrac{9}{2} c_{st} + 2 c_{\psi}, \nonumber
\\
\tilde{c}_{\delta, 3_{c_s}} & = & c_{\delta, 3_{c_s}} + \sfrac{1}{2 \pi} \sfrac{1}{c^2_{s(1)}} \sfrac{k_{NL}^2}{k_M^2} c_{\d^2 \delta, 1}, \nonumber
\\
\tilde{c}_{\delta^2, 1} & = & c_{\delta^2, 1} - \sfrac{17}{6} c_{s^2, 1}, \nonumber
\\
\tilde{c}_{\delta^2, 2} & = & c_{\delta^2, 2} - \sfrac{137}{16} c_{s^3} - \sfrac{71}{24} c_{st} - \sfrac{55}{42} c_{\psi} + \sfrac{7}{4} c_{\delta s^2}, \nonumber
\\
\tilde{c}_{s^2, 2} & = & c_{s^2, 2} - \sfrac{3}{4} c_{s^3} - \sfrac{1}{2} c_{st} - \sfrac{2}{7} c_{\psi}, \nonumber
\\
\tilde{c}_{\delta^3} & = & c_{\delta^3} + \sfrac{511}{72} c_{s^3} + \sfrac{25}{12} c_{st} + c_{\psi} - \sfrac{17}{6} c_{\delta s^2}.
\eea

\newpage

\section{Plots of bias parameters intervals and of Power Spectra}
\label{app:biasplot}

\begin{figure*}[h!]
   \begin{center}
   \hspace*{-0.5cm}
   \includegraphics[scale=0.55]{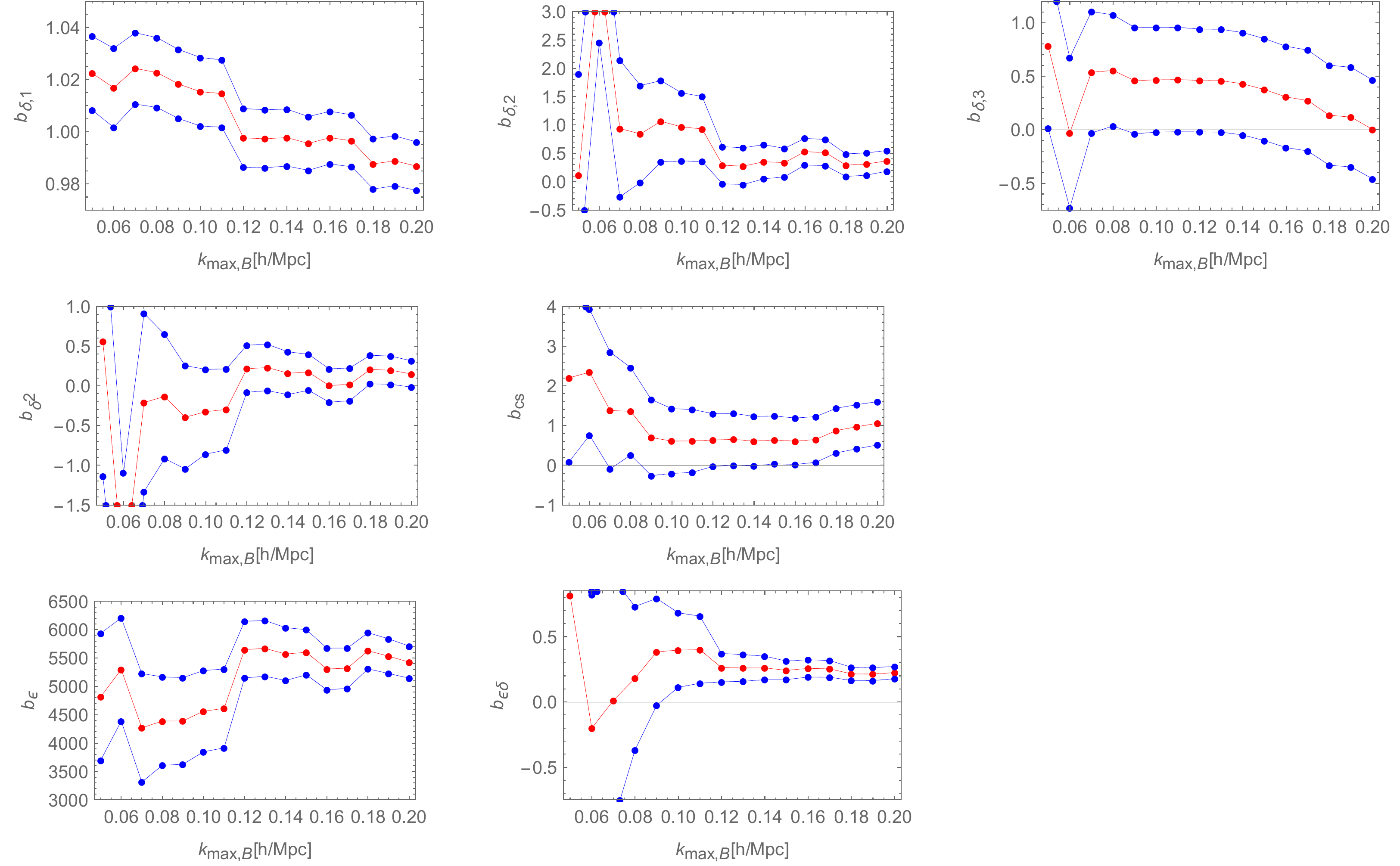}
   \end{center}
   \vspace*{-0.5cm}
   \caption{\small Plots of the values of the bias coefficients for Bin0. The points at $k<0.08 \, h \, {\rm Mpc}^{-1}$ present some unexpected behavior which is hard to justify given the fact thart the EFTofLSS is expected to work well at low wavenumber. We therefore ignore those points for the procedure to fix the values of the coefficients, described in Sec.~\ref{procedure}. Following this procedure, we find $k_{\rm fit} = 0.15 \, h \, {\rm Mpc}^{-1}$.}
   \label{fig:Bin0plotcoefficients}
\end{figure*}

\begin{figure*}[h!]
   \begin{center}
   \hspace*{-0.5cm}
   \includegraphics[scale=0.55]{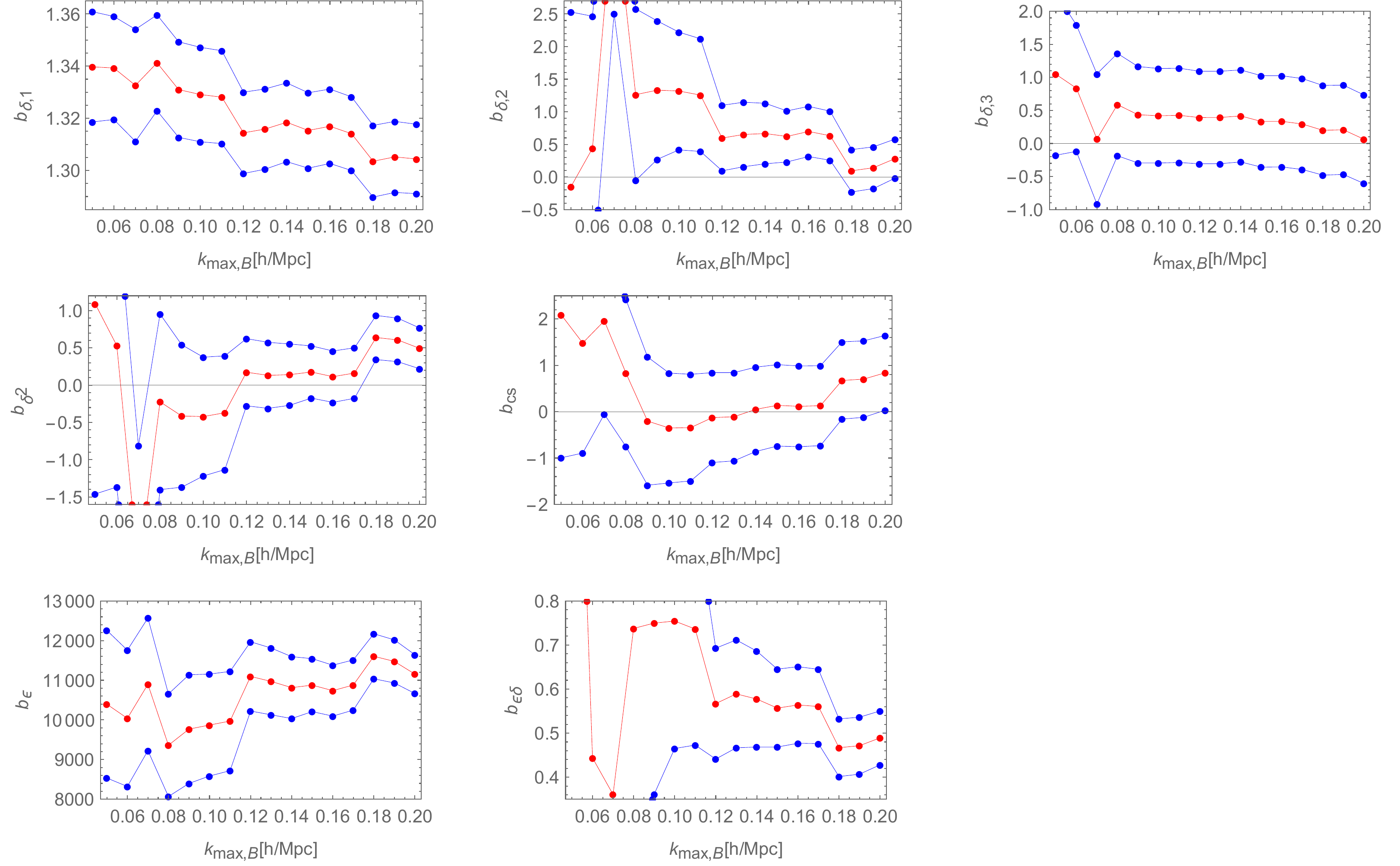}
   \end{center}
   \vspace*{-0.5cm}
   \caption{\small Plots of the values of the bias coefficients for Bin1. The points at $k<0.08 \, h \, {\rm Mpc}^{-1}$ present some unexpected behavior which is hard to justify given the fact that the EFTofLSS is expected to work well at low wavenumber. We therefore ignore those points for the procedure to fix the values of the coefficients, described in Sec.~\ref{procedure}. Following this procedure, we find $k_{\rm fit} = 0.18 \, h \, {\rm Mpc}^{-1}$.}
   \label{fig:Bin1plotcoefficients}
\end{figure*}

\begin{figure*}[h!]
   \begin{center}
   \hspace*{-0.5cm}
   \includegraphics[scale=0.49]{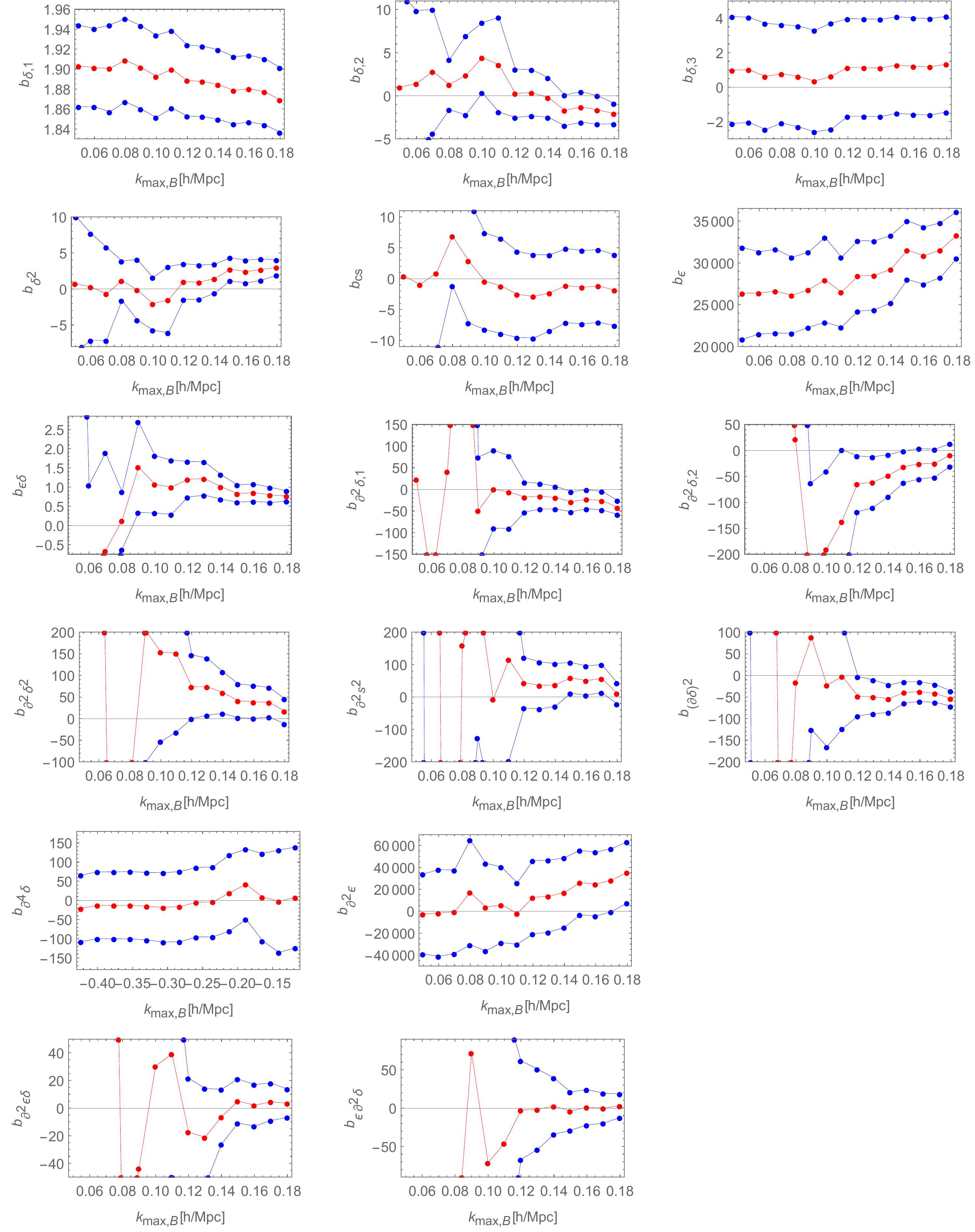}
   \end{center}
   \vspace*{-0.5cm}
   \caption{\small Plots of the values of the bias coefficients for Bin2. The points at $k<0.08 \, h \, {\rm Mpc}^{-1}$ present some unexpected behavior which is hard to justify given the fact that the EFTofLSS is expected to work well at low wavenumber. We therefore ignore those points for the procedure to fix the values of the coefficients, described in Sec.~\ref{procedure}. Following this procedure, we find $k_{\rm fit} = 0.15 \, h \, {\rm Mpc}^{-1}$.}
   \label{fig:Bin2plotcoefficients}
\end{figure*}

\begin{figure*}[h!]
   \begin{center}
   \hspace*{-0.5cm}
   \includegraphics[scale=0.55]{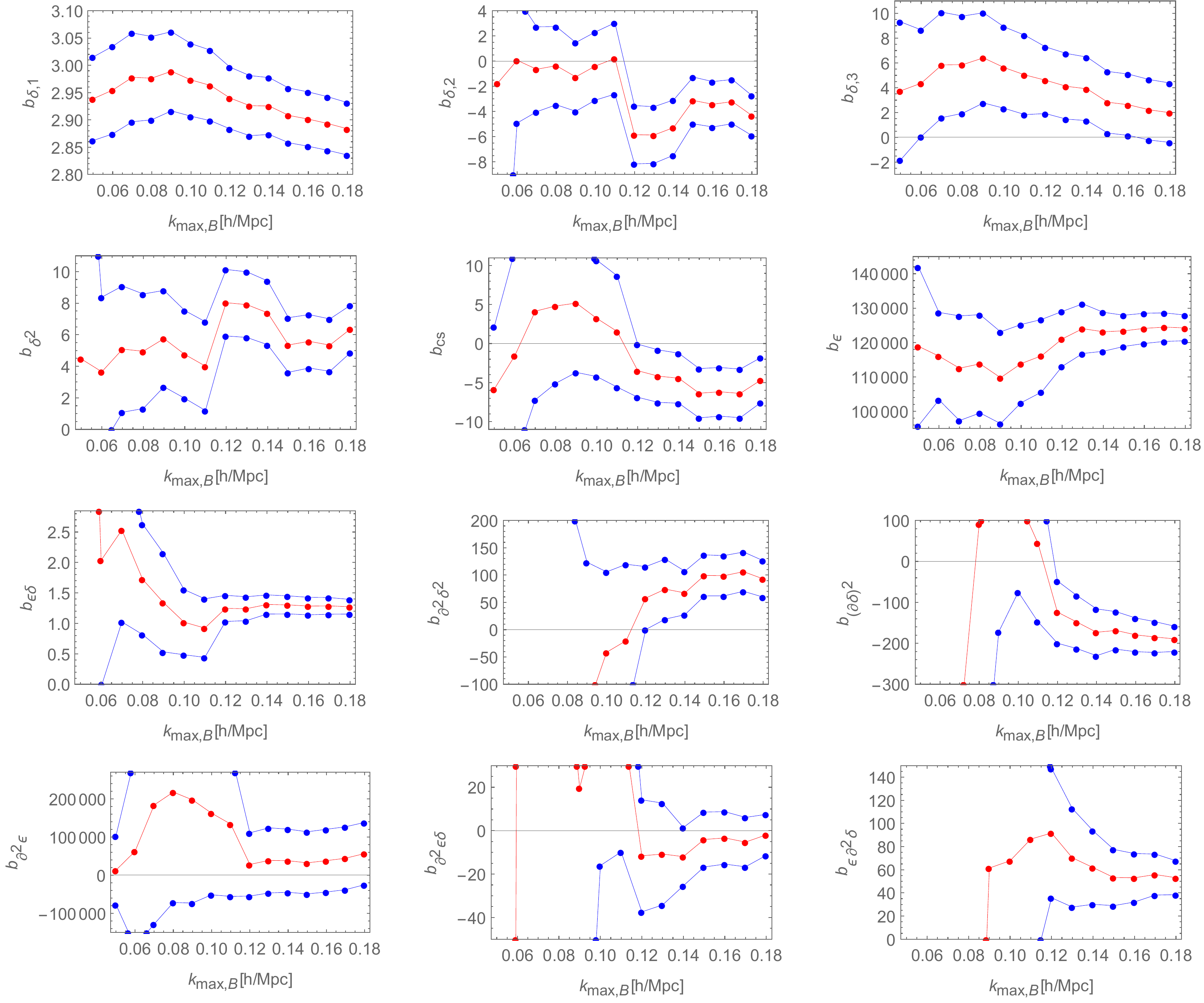}
   \end{center}
   \vspace*{-0.5cm}
   \caption{\small Plots of the values of the bias coefficients for Bin3. The points at $k<0.08 \, h \, {\rm Mpc}^{-1}$ present some unexpected behavior which is hard to justify given the fact that the EFTofLSS is expected to work well at low wavenumber. We therefore ignore those points for the procedure to fix the values of the coefficients, described in Sec.~\ref{procedure}. Following this procedure, we find $k_{\rm fit} = 0.12 \, h \, {\rm Mpc}^{-1}$.}
   \label{fig:Bin3plotcoefficients}
\end{figure*}

In this Appendix we present the plots of the best fit parameters as a function of $k_{\rm max,B}$ for each bin. We add the 2-$\sigma$ error bars to each plot. Following the procedure of Sec.~\ref{procedure}, we find the following $k_{\rm fit}$'s :
\begin{itemize}
\item For Bin0, Fig.~\ref{fig:Bin0plotcoefficients}, $k_{\rm fit} = 0.15 \, h \, {\rm Mpc}^{-1}$,
\item For Bin1, Fig.~\ref{fig:Bin1plotcoefficients}, $k_{\rm fit} = 0.18 \, h \, {\rm Mpc}^{-1}$,
\item For Bin2, Fig.~\ref{fig:Bin2plotcoefficients}, $k_{\rm fit} = 0.15 \, h \, {\rm Mpc}^{-1}$,
\item For Bin3, Fig.~\ref{fig:Bin3plotcoefficients}, $k_{\rm fit} = 0.12 \, h \, {\rm Mpc}^{-1}$.
\end{itemize}

\begin{figure*}[h!]
   \begin{center}
   \hspace*{-0.5cm}
   \includegraphics[scale=0.37]{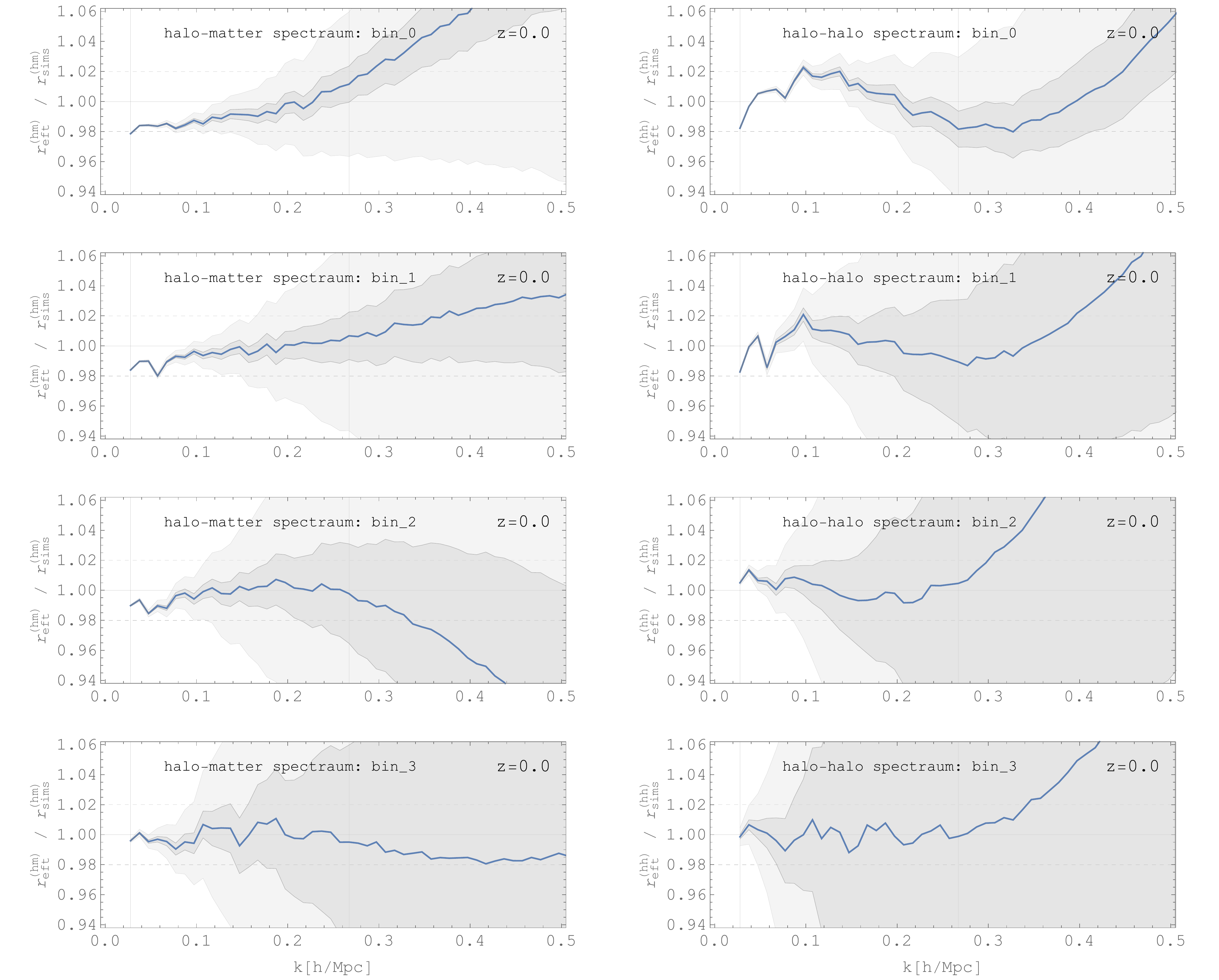}
   \end{center}
   \vspace*{-0.5cm}
   \caption{\small {Comparisons between the theoretical prediction and the numerical simulations of the cross power spectrum, on the left-hand side, and the auto power spectrum, on the right-hand side. The blue line represents the division of the theoretical prediction of the ratio between the cross or auto power spectrum and the dark matter power spectrum and the same quantity but measured in numerical simulations. The grey areas are order-of-magnitude estimates of the error, following the formulas in \cite{Angulo:2015eqa}.}}
   \label{fig:powerspectraplots}
\end{figure*}

Note that when we use this procedure, we discard the points with $k<0.08 \, h \, {\rm Mpc}^{-1}$ because they appear to have some unjustified behaviour in a region where the theory is expected to work reasonably well.  In Fig.~\ref{fig:powerspectraplots}, we also present the plots of the two-point functions predictions versus the numerical data.

\section{Correlation matrices of the bias parameters}
\label{app:biascorrelation}

In this Appendix we present the correlation matrices for the bias coefficients as obtained with our fitting procedure.

\begin{table*}[h!]
\caption{{Correlation matrix for the bias coefficients for Bin0.}}
\centering 
\setlength{\tabcolsep}{8pt}
\renewcommand{\arraystretch}{1.0}
\begin{tabular}{c|ccccccc}
\hline\hline
Bin0 & $b_{\delta,1}$ & $b_{\delta,2}$ & $b_{\delta,3}$ & $b_{\delta^2}$ & $b_{cs}$ & $b_{\epsilon}$ & $b_{\epsilon \delta}$ \\
\hline
$b_{\delta, 1}$ & 1. & 0.08 & 0.69, & -0.05 & -0.481 & 0.03 & -0.11 \\
\hline
$b_{\delta, 2}$ & 0.08 & 1. & -0.26, & -0.99 & -0.15 & -0.86 & 0.84 \\
\hline
$b_{\delta, 3}$ & 0.69 & -0.26 & 1., & 0.25 & -0.80 & 0.43 & -0.28 \\
\hline
$b_{\delta^2}$ & -0.05 & -0.99 & 0.25 & 1. & 0.21, & 0.80 & -0.87 \\
\hline
$b_{cs}$ & -0.48 & -0.15 & -0.80 & 0.21 & 1. & -0.22 & -0.24 \\
\hline
$b_{\epsilon}$ & 0.02 & -0.86 & 0.43 & 0.80 & -0.22 & 1. & -0.60 \\
\hline
$b_{\epsilon \delta}$ & -0.11 & 0.84 & -0.28 & -0.87 & -0.24 & -0.60 & 1. \\
\hline
\end{tabular}
\label{tb:corrmatrix0}
\end{table*}

\begin{table*}[h!]
\caption{{Correlation matrix for the bias coefficients for Bin1.}}
\centering 
\setlength{\tabcolsep}{8pt}
\renewcommand{\arraystretch}{1.0}
\begin{tabular}{c|ccccccc}
\hline\hline
Bin1 & $b_{\delta,1}$ & $b_{\delta,2}$ & $b_{\delta,3}$ & $b_{\delta^2}$ & $b_{cs}$ & $b_{\epsilon}$ & $b_{\epsilon \delta}$ \\
\hline
$b_{\delta, 1}$ & 1. & -0.05 & 0.71, & 0.07 & -0.52 & 0.17 & -0.29 \\
\hline
$b_{\delta, 2}$ & -0.05 & 1. & -0.27, & -0.99 & -0.16 & -0.82 & 0.88 \\
\hline
$b_{\delta, 3}$ & 0.71 & -0.27 & 1., & 0.27 & -0.81 & 0.43 & -0.43 \\
\hline
$b_{\delta^2}$ & 0.07 & -0.99 & 0.27 & 1. & 0.20, & 0.76 & -0.89 \\
\hline
$b_{cs}$ & -0.52 & -0.16 & -0.81 & 0.20 & 1. & -0.21 & -0.05 \\
\hline
$b_{\epsilon}$ & 0.17 & -0.82 & 0.43 & 0.76 & -0.21 & 1. & -0.75 \\
\hline
$b_{\epsilon \delta}$ & -0.29 & 0.88 & -0.43 & -0.89 & -0.05 & -0.75 & 1. \\
\hline
\end{tabular}
\label{tb:corrmatrix1}
\end{table*}

\begin{table*}[h!]
\caption{{Correlation matrix for the bias coefficients for Bin2.}}
\centering 
\setlength{\tabcolsep}{2.7pt}
\renewcommand{\arraystretch}{1.0}
\begin{tabular}{c|cccccccccccccccc}
\hline\hline
Bin2 & $b_{\delta,1}$ & $b_{\delta,2}$ & $b_{\delta,3}$ & $b_{\delta^2}$ & $b_{cs}$ & $b_{\epsilon}$ & $b_{\epsilon \delta}$ & $b_{\partial^2 \delta,1}$ & $b_{\partial^2 \delta,2}$ & $b_{\partial^2 \delta^2}$ & $b_{\partial^2 s^2}$ & $b_{(\partial \delta)^2}$ & $b_{\partial^4 \delta}$ & $b_{\partial^2 \epsilon}$ & $b_{\partial^2 \epsilon \delta}$ & $b_{\epsilon \partial^2 \delta}$ \\
\hline
$b_{\delta, 1}$ & 1. & 0.28 & 0.65 & -0.23 & -0.40 & -0.43 & 0.05 & 0.12 & -0.31 & 0.18 & 0.05 & 0.26 & 0.40 & -0.23 & 0.10 & 0.11 \\
\hline
$b_{\delta, 2}$ & 0.28 & 1. & -0.002 & -0.99 & -0.39 & -0.76 & 0.81 & 0.60 & -0.45 & 0.56 & -0.15 & 0.01 & 0.16 & -0.42 & -0.01 & -0.48 \\
\hline
$b_{\delta, 3}$ & 0.65 & -0.002 & 1. & 0.003 &, -0.80 &, 0.10 & -0.12 & -0.16 & -0.12 & 0.10 & 0.05 & 0.03 & 0.85 & -0.04 & -0.003 & 0.05 \\
\hline
$b_{\delta^2}$ & -0.23 & -0.99 & 0.003 & 1. & 0.44 & 0.68 & -0.82 & -0.57 & 0.43 & -0.58 & 0.16 & 0.05 & -0.19 & 0.35 & 0.04 & 0.53 \\
\hline
$b_{cs}$ & -0.40 & -0.39 & -0.80 & 0.44 & 1. & -0.03 & -0.30 & -0.03 & 0.22 & -0.38 & 0.06 & 0.19 & -0.92 & -0.08 & 0.11 & 0.35 \\
\hline
$b_{\epsilon}$ & -0.43 & -0.76 & 0.10 & 0.68 & -0.03 & 1. & -0.55 & -0.63 & 0.39 & -0.29 & 0.07 & -0.31 &, 0.09 & 0.53 & -0.14 & 0.10 \\
\hline
$b_{\epsilon \delta}$ & 0.05 & 0.81 & -0.12 & -0.82 & -0.30 & -0.55 & 1. & 0.52 & -0.44 & 0.50 & -0.04 & -0.17 & 0.11 & -0.17 & -0.17 & -0.60 \\
\hline
$b_{\partial^2 \delta, 1}$ & 0.12 & 0.60 & -0.16 & -0.57 & -0.03 & -0.63 & 0.52 & 1. & -0.29 & 0.36 & -0.001 & 0.12 & -0.06 & -0.30 & -0.16 & -0.39 \\
\hline
$b_{\partial^2 \delta, 2}$ & -0.31 & -0.45 & -0.12 & 0.43 & 0.22 & 0.39 & -0.44 & -0.29 & 1. & -0.88 & -0.74 & 0.15 & -0.13 & 0.23 & 0.18 & 0.13 \\
\hline
$b_{\partial^2 \delta^2}$ & 0.18 & 0.56 & 0.10 & -0.58 & -0.38 & -0.29 & 0.50 & 0.36 & -0.88 & 1. & 0.58 & -0.09 & 0.22 & -0.11 & -0.09 & -0.47 \\
\hline
$b_{\partial^2 s^2}$ & 0.05 & -0.15 & 0.05 & 0.16 & 0.06 & 0.07 & -0.04 & -0.001 & -0.74 & 0.58 & 1. & -0.13 & -0.01 & 0.04 & -0.15 & 0.11 \\
\hline
$b_{(\partial \delta)^2}$ & 0.26 & 0.01 &, 0.03 & 0.05 & 0.19 & -0.31 & -0.17 & 0.12 & 0.15 & -0.09 & -0.13 & 1. & -0.12 & -0.29 & 0.66 & -0.09 \\
\hline
$b_{\partial^4 \delta}$ & 0.40 & 0.16 & 0.85 & -0.19 & -0.92 & 0.09 & 0.11 & -0.06 & -0.13 & 0.22 & -0.01 & -0.12 & 1. & 0.15 & -0.07 & -0.18 \\
\hline
$b_{\partial^2 \epsilon}$ & -0.23 & -0.42 & -0.04 & 0.35 & -0.08 & 0.53 & -0.17 & -0.30 & 0.23 & -0.11 & 0.04 & -0.29 & 0.15 & 1. & -0.12 & -0.07 \\
\hline
$b_{\partial^2 \epsilon \delta}$ & 0.10 & -0.01 & -0.003 & 0.04 & 0.11 & -0.14 & -0.17 & -0.16 & 0.18 & -0.09 & -0.15 & 0.66 & -0.07 & -0.12 & 1. & -0.32 \\
\hline
$b_{\epsilon \partial^2 \delta}$ & 0.11 & -0.48 & 0.05 & 0.53 & 0.35 & 0.10 & -0.60 & -0.39 & 0.13 & -0.47 & 0.11 & -0.09 & -0.18 & -0.07 & -0.32 & 1. \\
\hline
\end{tabular}
\label{tb:corrmatrix2}
\end{table*}

\begin{table*}[h!]
\caption{{Correlation matrix for the bias coefficients for Bin3.}}
\centering 
\setlength{\tabcolsep}{6pt}
\renewcommand{\arraystretch}{1.0}
\begin{tabular}{c|cccccccccccc}
\hline\hline
Bin3 & $b_{\delta,1}$ & $b_{\delta,2}$ & $b_{\delta,3}$ & $b_{\delta^2}$ & $b_{cs}$ & $b_{\epsilon}$ & $b_{\epsilon \delta}$ & $b_{\partial^2 \delta^2}$ & $b_{(\partial \delta)^2}$ & $b_{\partial^2 \epsilon}$ & $b_{\partial^2 \epsilon \delta}$ & $b_{\epsilon \partial^2 \delta}$ \\
\hline
$b_{\delta, 1}$ & 1. & -0.22 & 0.82 & 0.34 & 0.38 & -0.62 & -0.31 & -0.49 & 0.37 & 0.04 & 0.29 & 0.32 \\
\hline
$b_{\delta, 2}$ & -0.22 & 1. & -0.47 & -0.98 & -0.44 & -0.18 & 0.29 & 0.29 & 0.09 & -0.34 & 0.17 & -0.18 \\
\hline
$b_{\delta, 3}$ & 0.82 & -0.47 & 1. & 0.58 & 0.35 & -0.48 & -0.36 & -0.54 & 0.31 & -0.09 & 0.22 & 0.34 \\
\hline
$b_{\delta^2}$ & 0.34 & -0.98 & 0.58 & 1. & 0.56 & 0.01 & -0.32 & -0.40 & -0.0004 & 0.29 & -0.08 & 0.24 \\
\hline
$b_{cs}$ & 0.38 & -0.44 & 0.35 & 0.56 & 1. & -0.63 & -0.16 & -0.61 & 0.27 & 0.12 & 0.27 & 0.32 \\
\hline
$b_{\epsilon}$ & -0.62 & -0.18 & -0.48 & 0.01 & -0.63 & 1. & 0.06 & 0.58 & -0.46 & 0.05 & -0.50 & -0.31 \\
\hline
$b_{\epsilon \delta}$ & -0.31 & 0.29 & -0.36 & -0.32 & -0.16 & 0.06 & 1. & 0.03 & -0.13 & 0.04 & -0.27 & -0.53 \\
\hline
$b_{\partial^2 \delta^2}$ & -0.49 & 0.29 & -0.54 & -0.40 & -0.61 & 0.58 & 0.03 & 1. & 0.09 & -0.005 & -0.02 & -0.56 \\
\hline
$b_{(\partial \delta)^2}$ & 0.37 & 0.09 & 0.31 & -0.0004 & 0.27 & -0.46 & -0.13 & 0.09 &, 1. & -0.13 & 0.61 & -0.14 \\
\hline
$b_{\partial^2 \epsilon}$ & 0.04 & -0.34 & -0.09 & 0.29 & 0.12 & 0.05 & 0.04 & -0.005 & -0.13 & 1. & -0.09 & -0.01 \\
\hline
$b_{\partial^2 \epsilon \delta}$ & 0.29 & 0.17 & 0.22 & -0.08 & 0.27 & -0.50 & -0.27 & -0.02 & 0.61 & -0.09 & 1. & -0.31 \\
\hline
$b_{\epsilon \partial^2 \delta}$ & 0.32 & -0.18 & 0.34 & 0.24 & 0.32 & -0.31 & -0.53 & -0.56 & -0.14 & -0.01 & -0.31 & 1. \\
\hline
\end{tabular}
\label{tb:corrmatrix3}
\end{table*}

\newpage

\vfill

\bibliographystyle{JHEP}


\end{document}